\documentclass{aa}  
\usepackage[utf8]{inputenc}
\usepackage{slashbox}
\usepackage{natbib}
\usepackage[colorlinks]{hyperref}
\usepackage{xcolor}
\usepackage{ulem}
\usepackage{booktabs}
\usepackage{array}
\usepackage{adjustbox}

\hypersetup{
    citecolor=[RGB]{44, 117, 255},  
}
\usepackage{amsmath}
\usepackage{listings,comment}

\def\nclu{138}
\def\nrad{50}
\def\tng{\texttt{IllustrisTNG}}

\def\Rmin{$R_{\text{min}}$}
\def\Rmax{$R_{\text{max}}$}
\def\RminEq{R_{\text{min}}}
\def\RmaxEq{R_{\text{max}}}
\def\rhomin{$\rho_{\text{min}}$}
\def\rhomax{$\rho_{\text{max}}$}
\def\rhominEq{\rho_{\text{min}}}
\def\rhomaxEq{\rho_{\text{max}}}

\def\ovi{O~VI}
\def\ovii{O~VII}
\def\oviii{O~VIII}

\def\neix{Ne~IX}
\def\nex{Ne~X}

\def\nodata{...}
\usepackage{nicefrac}

\makeatletter
\renewcommand*\aa@pageof{, page \thepage{} of \pageref*{LastPage}}
\makeatother

\title{Soft X-ray emission from warm gas in IllustrisTNG circum--cluster environments} 
\subtitle{I -- Sample Properties, Methods and Initial Results}
\titlerunning{}

\author{
  Celine~Gouin\inst{1}\thanks{E-mail:~\tt{celine.gouin@ias.u-psud.fr}},  
  Massimiliano~Bonamente\inst{2},  
  Daniela~Gal\'arraga-Espinosa\inst{3}, 
  Stephen~Walker\inst{2}, 
  Mohammad~Mirakhor \inst{2}
}

\institute{
Université Paris-Saclay, CNRS, Institut d’Astrophysique Spatiale, 91405, Orsay, France
\label{inst1}
\and
University of Alabama in Huntsville, Department of Physics and Astronomy, Huntsville, AL 35899
\label{inst3}
\and
Max-Planck Institute for Astrophysics, Karl-Schwarzschild-Str. 1, D-85741 Garching, Germany
\label{inst2}
}

\date{\today}

\abstract
{
\textit{Context.} Whereas X-ray clusters are extensively used for cosmology, their idealistic modelling, through the hypotheses of spherical symmetry and hydrostatic equilibrium, are more and more being questioned. Along these lines, the soft X-ray emission detected in tens of clusters with {ROSAT} was found to be higher than what expected from the idealistic hot gas modelling, pointing to our incomplete understanding of these objects.\\ 
\textit{Aims.} Given that cluster environments are at the interface between the hot intra--cluster medium (ICM), warm circumgalactic medium (WCGM) and warm--hot intergalactic medium (WHIM), we aim to explore the relative soft X-ray emission of different gas phases in circum–cluster environments. \\ 
\textit{Method.}  By using the most massive halos in IllustrisTNG at $z=0$, we have predicted the hydrodynamical properties of the gas from cluster centers to their outskirts ($5 \ R_{200}$), and modelled their X–ray radiation for various plasma phases. \\
\textit{Results.} First, we found that the radial profile of temperature, density, metallicity and clumpiness of the ICM are in good agreement with recent X-ray observations of clusters. 
Secondly, we have developed a method to predict the radial profile of soft X-ray emission in different bands, the column density of ions and the X–ray absorption lines (\oviii, \ovii, \neix, and \neix) of warm-hot gas inside and around clusters.  \\
\textit{Conclusion.} The warm gas (in the form of both WCGM and WHIM gas) is a strong emitter in soft X-ray bands, and is qualitatively consistent with the observational measurements. Our results suggest that the cluster \textit{soft excess} is induced by the thermal emission of warm gas in the circum–cluster environments.
}


\authorrunning{Gouin et al.}

\begin{document}
\maketitle

\section{Introduction: X--rays from circum--cluster environments and the missing baryons problem}

X--ray emission from galaxy clusters is used extensively to 
probe a variety of astrophysical phenomena and to study cosmology.
Since the early surveys of the X--ray sky \citep{giacconi1972}, it became apparent that clusters
are powerful sources of X--ray emission, originating primarily from a hot phase ($\geq 10^7$~K) of
the intergalactic medium that is kept bound by a massive dark matter halo that significantly outweighs
visible galaxies \citep[e.g.][]{bahcall1977}. 
The X--ray emission is primarily thermal in origin, with free--free bremsstrahlung providing the bulk of the
emissivity, with the presence of emission lines from such elements as iron that are not
fully ionized \citep[e.g.][]{mitchell1976,serlemitsos1977}.

 Among some of the key uses of X--ray emission from galaxy clusters are the determination of cosmological parameters that describe dark matter and dark energy \citep[e.g.][]{allen2004,vikhlinin2009,mantz2014, mantz2022} and the Hubble constant 
 \citep[e.g.][]{bonamente2006,wan2021},
the understanding of the growth of cosmic structure on the largest scales \citep[see][ for recent review]{walker2019}, and the study of the complex feedback mechanism between the cluster's galaxies and their active galactic nuclei \citep[e.g.][]{Yang2016}.
At the heart of all these studies is the interpretation of the detected emission as radiation from
a hot intra--cluster medium (ICM) that is in approximate hydrostatic equilibrium with its underlying
matter potential.

At soft X--ray wavelengths, broadly defined as photon energies substantially below the
peak of the X--ray emission from the hot ICM ($\leq 1$~keV) and above the effective cut--off imposed by Galactic absorption ($\sim 0.1$~keV), X-ray emission from and near clusters
is less well studied. This is due to a combination of the lower effective
energy and poorer calibration of most X--ray missions \citep[e.g.][]{nevalainen2010}
towards lower photon energies, the effects of Galactic foregrounds \citep[e.g.][]{snowden1995} that are both time-- and space--variables, and the contamination from
sub--virial phases.

Numerical simulations provide the means to study the thermodynamics of the gas in and near cluster environments. With the advanced of baryonic physic models through sub-grid and particle techniques, the current generation of large hydrodynamical simulations, such as \tng\ \citep{Nelson2019}, \texttt{HorizonAGN} \citep{Dubois2016}, \texttt{Magneticum} \citep{Hirschmann2014, Dolag2016, Ragagnin2017MAGNETICUM} and \texttt{TheThreeHundred} project \citep{Cui2018}, provides statistical sample of highly resolved simulated clusters to deeply investigate their gas properties, and in particular their thermal footprint for X-Ray interpretation. Probing the gas physics through state-of-the-art simulations have already revealed departure to hydrostatic equilibrium inside clusters \citep[e.g.][]{Gianfagna2023}, resulting of turbulence and bulk motion \citep[e.g.][]{Angelinelli2020}, and proposed correction for this bias in X-ray  \citep{Ansarifard2020}. More generally, numerous investigations have been recently done to probe the ICM physics by using numerical simulations, such as their chemical enrichment \citep{Biffi2018}, and their thermal and pressure profiles for accurately predict scaling relations \citep[e.g.][]{Barnes2017,Pop2022} to better interpret X-ray emission of clusters. 

For our cluster gas investigation, we are using the TNG300-1 simulation, the larger box of \tng\ simulations ($\sim 300\ \mathrm{Mpc}^3$) with the highest spatial resolution, to take advantage of both, a statistical sample of clusters, and an accurate modeling of the baryonic physics \citep{Pillepich2018}. 
Using TNG300-1 simulation, \cite{Martizzi19} and \cite{Galarraga21} have highlighted that hot gas is dominant inside clusters, as the knots of the large-scale cosmic web, whereas warm diffuse gas can be used to trace filamentary structures at  $z=0$. Such improvement in our understanding of cluster and cosmic gas has recently allow to forecast on warm and hot gas observations through OVII and OVIII emission lines for the next generation spectrometers \citep{Parimbelli22,butler2023,Tuominen2023}. 
Focusing exclusively on cluster environments, \cite{gouin2022} have explored the gas thermodynamics in TNG300-1, showing that clusters are the interface between two dominated gas phases: the warm and the hot gas. The warm gas in filaments infalls to the connected clusters, undergoes motions of accretion and ejection at cluster borders, and transitions then into the hot gas phase, inside clusters.

This project aims to fill the gap in our theoretical and observational understanding
of soft X--ray emission in and near galaxy clusters, using the \cite{gouin2022} \tng\ sample of massive clusters
studied out to $5\times r_{200}$. We refer to these astrophysical volumes as the \textit{circum--cluster environment} \citep[as introduced by][]{yoon2013}, i.e, 
the region in and surrounding galaxy clusters, including the transition between the inner virialized ICM and 
the outer accreting sub--virial plasmas that is expected to contain the highest--density WHIM.
The main motivation for this study of the soft X--ray emission near clusters is the
expectation that a significant fraction of the universe's baryons at the present epoch
is in a warm--hot intergalactic medium (WHIM) at $\log T(\text{K})\leq 7$ \citep[e.g.][]{cen1999,dave2001}. These baryons are widely expected to bridge the \textit{missing
baryons} gap, i.e., resolve the observational discrepancy between the lower amount 
of low--$z$ baryons compared to high--$z$ baryons \citep[e.g.][]{danforth2016}.
Although evidence is mounting that WHIM filaments are responsible for 
both emission \citep[e.g.][]{Tanimura2020,mirakhor2022,Tanimura2022} and absorption of X--rays \citep[e.g.][]{kovacs2019,spence2023} in amounts that are consistent with with the missing fraction of baryons, 
the regions near clusters where filaments are expected to be densest remain relatively unexplored, in part because of the difficulties on disentangling the contributions from the virial and sub--viral phases. One exception is the detection of the cluster \textit{soft excess} emission
in a number of clusters with {ROSAT} \citep[e.g.][]{bonamente2003, bonamente2022c} 
and other soft X--ray instruments, which is consistent with the detection of WHIM filaments
projecting onto massive clusters.
One of the main motivations of this project is therefore to address the contribution to the projected X--ray emission
at soft X--ray energies from hot and sub--virial phases in and around massive clusters, and other aspects
of cluster physics that are related to this X--ray emission.

This paper, which is intended to be the first of a series, describes the methods of 
analysis of the \tng\ data to obtain averaged radial profiles of thermodynamic quantities of interest for the prediction of the X--ray emission, via density, temperature, chemical abundances, volume covering fraction and clumpiness of the gas phases, and initial results on the projection of the emissivity from the different gas phases in and around clusters.
This paper is structured as follows: Sect.~\ref{sec:TNG} presents the \tng\ simulations and their analysis to obtain radial profiles; Sect.~\ref{sec:projection} describes our analytical method to project radial profiles on a detector plane,
Sect.~\ref{sec:SX} describes the methods to obtain projected profiles of
radiated power and X--ray surface brightness that can be compared with
observations, and preliminary results from that analysis.
Sect.~\ref{sec:conclusions} presents a brief discussion and our conclusions.
Subsequent papers will make use of these methods for a more in--depth analysis of the emission from sub--virial gas in and around clusters, its astrophysical and cosmological implications, and related aspects of cluster physics.

\section{Thermodynamical profiles from \tng\ circum--cluster environments: Methods and comparison with observations}
\label{sec:TNG}

\subsection{Cluster selection from IllustrisTNG simulation}

\begin{table*}[]
    \centering
    \begin{tabular}{c| l| l| l | l}
         Sample & Mass Range & $\langle M_{200}\rangle$  & $\langle R_{200}\rangle$ & Number of clusters \\
        \hline 
         All clusters & $\log( M_{200}[M_{\odot}/h])  >14$ & $1.97 \times 10^{14} M_{\odot}/h$ & 0.92/$h$ Mpc& 138 \\[5pt]
         $M_1$ & $14.0 <\log( M_{200}[M_{\odot}/h]) <14.1$ & $1.13 \times 10^{14} M_{\odot}/h$ & 0.78/$h$ Mpc & 47\\[5pt]
         $M_2$ & $14.1 <\log( M_{200}[M_{\odot}/h]) <14.5$& $1.91 \times 10^{14} M_{\odot}/h$ & 0.92/$h$ Mpc& 78\\[5pt]
         $M_3$ & $ \log( M_{200}[M_{\odot}/h]) >14.5$ & $5.37 \times 10^{14} M_{\odot}/h$ & 1.3/$h$ Mpc & 13\\
    \hline
    \end{tabular}
    \caption{Main properties of the all cluster sample, and it three mass bins $M_1$, $M_2$ and $M_3$.}
    \label{tab:sample}
\end{table*}

Starting from the halo catalog of the TNG300-1 simulation at $z=0$, we select all friends-of-friends halos (FoF) with masses $M_{200}>10^{14} M_{\odot}/h$. 
In this catalog, the radial scale of FoF halos $R_{200}$ is defined as the radius of a sphere which encloses a mass $M_{200}$ with an average density equal to 200 times the critical density of the Universe.
From our mass selection of 153 clusters, we consider only the 138 clusters whose centers
are more distant than $5 \times R_{200}$ from the simulation box edges, to fully explore cluster environments.
 The main properties of the \tng\ cluster sample are reported in Table~\ref{tab:sample}. In this table, we define also the three mass bins, considering in the following analysis, the low-mass clusters $M_1$ (with $M_{200} \in 10^{[14.0-14.1]} M_{\odot}/h$), the middle cluster mass range $M_2$ (with $M_{200} \in 10^{[14.1-14.5]} M_{\odot}/h$), and the most massive clusters $M_3$ with $M_{200} > 10^{14.5} M_{\odot}/h$.

\subsection{Gas phases}
\begin{figure}
    \centering
    \includegraphics[width=0.45\textwidth]{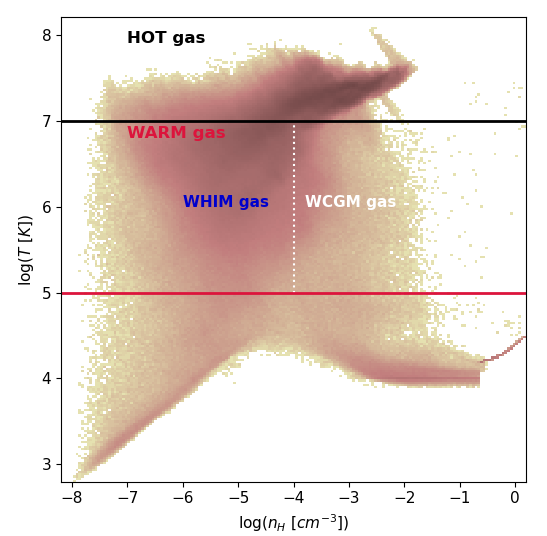}
    \caption{Phase-space of the gas cells around a given cluster in the TNG300-1 simulation, shown by the normalised 2D histogram. The hot gas is defined with a temperature $T[\mathrm{K}] >10^7 $, and the WARM gas with temperature between $10^5<T [\mathrm{K}]<10^7$. The WHIM gas is a sub-sample of the WARM gas, with the condition to be diffuse, i.e., its density is $n_H [\mathrm{cm}^{-3}]< 10^{-4}$. }
    \label{fig:Illustration_gas_phase}
\end{figure}
\begin{figure*}
    \centering
    \includegraphics[width=0.95\textwidth]{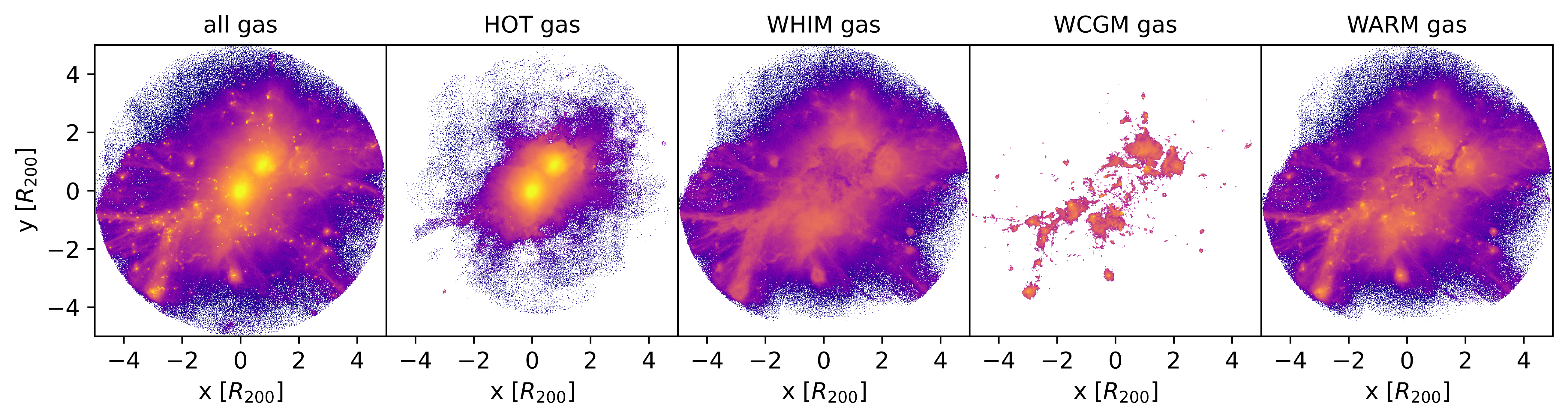}
    \caption{Distribution of gas within $5 \ R_{200}$ around a given massive cluster. The first panel shows the distribution of all the gas cells, the second one only the HOT gas, the third one only the WHIM gas, and the left panel shows the WARM gas.}
    \label{fig:Illustration_gas_dist}
\end{figure*}

The gas cells are separated depending of their temperature and density, in order to distinguish different gas phases. We have followed the temperature and density cuts of \cite{Martizzi19}, commonly used in hydrodynamical simulation analyses to explore gas phase transitions \citep[see e.g.][]{Galarraga21}. As demonstrated by \cite{gouin2022}, cluster environments are dominated by two main phases, mostly the Hot ICM ($T [\mathrm{K}]>10^7$) inside the clusters, and the warm diffuse gas phase (WHIM: $10^5< T [\mathrm{K}]<10^7$ and $n_H[\mathrm{cm}^{-3}]<10^{-4}$) outside clusters \citep[see Fig. 1 and 2 of][]{gouin2022}. In this study, we defined also the WARM gas as all gas cells with warm temperature $10^5< T[\mathrm{K}] <10^7$, thus accounting of both WHIM and WCGM gas phase ($10^5< T [\mathrm{K}]<10^7$ and $n_H[\mathrm{cm}^{-3}]>10^{-4}$). We illustrated this different gas phase definitions in the phase-space diagram Fig. \ref{fig:Illustration_gas_phase}. Although the distinction between WHIM and WCGM is used as a simple means to identify the lower-- and higher--density portions of the WARM sub--virial phase.

In Fig. \ref{fig:Illustration_gas_dist}, we show the spatial distribution of these different gas phases for a given cluster environment (up to $5 \ R_{200}$).
As expected from \cite{gouin2022}, the hot gas phase is tracing the gas distribution inside clusters, whereas WHIM gas is representing the filamentary patterns around cluster \citep[see also][]{Planelles2018}. Interestingly, our extend definition of WARM gas, show the WHIM gas with the addition of warm clumps gas. Indeed, by plotting only WCGM gas in Fig. \ref{fig:Illustration_gas_dist}, one can see that small dense clumps of gas are resulting from WCGM phase. As expected by \cite{Martizzi19}, this definition of WCGM and WHIM gases therefore represents physically different gas phases.

\subsection{Numerical computation of radial thermodynamic profiles}

\label{sec:profiles}
\begin{figure}
    \centering
    \includegraphics[width=3.5in]{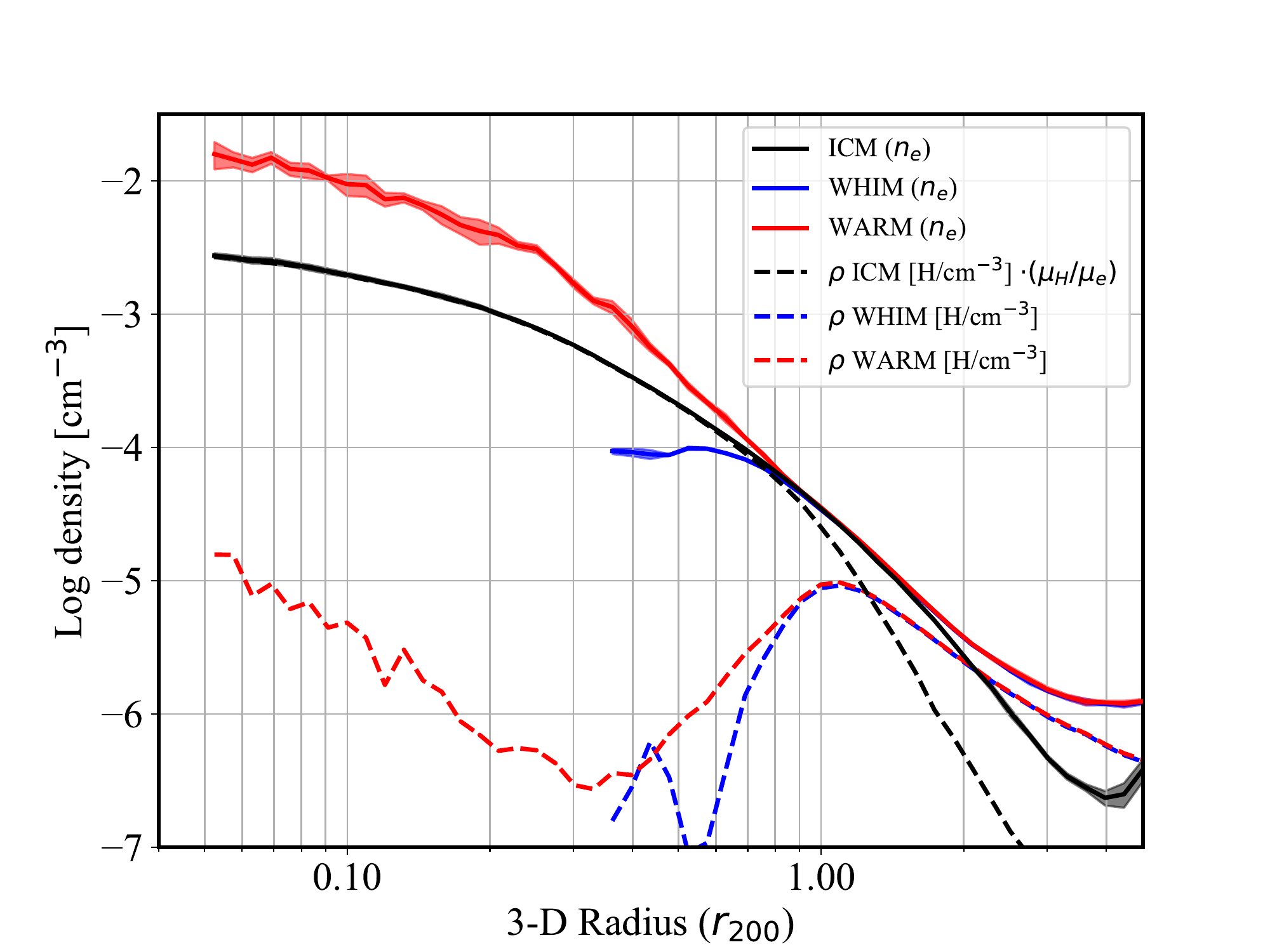}
    \caption{Right panel : The electron density $n_e$ calculated according to the averaging of Eq.~\ref{eq:Qaverage} is the solid line, and the mass density calculated according to the averaging of Eq.~\ref{eq:rho} is in dashed line. We show the median profile over all the cluster sample.}
    \label{fig:ElDens}
\end{figure}

For each thermodynamic quantity of the gas, we average the value by considering the gas cells (labeled by the index $i$ and with a volume $V_i$) inside radial shells (labeled by the index $k$, with $r_k$ normalized by $R_{200}$) around the cluster center. The average of a quantity $Q$ is given by:
\begin{equation}
    Q(r_{k}) = \dfrac{\sum\limits_{i \in R_k} Q_i \times V_i}{\sum\limits_{i \in R_k} V_i}
    \label{eq:Qaverage}
\end{equation}
with the $k=1,..,50$ index labelling the 50 logarithmically--spaced radial distances around the center, ranging from $0$ to $5 \ R_{200}$, and the radial range is $R_k=[r_{k-1},r_{k}]$. 
The averages according to \eqref{eq:Qaverage} therefore represent the volume-weighted average value of the quantity for a given gas phase (either all gas cells, HOT, WHIM, or WARM gas). 
These averages are shown as solid lines in Figures~\ref{fig:ElDens}, \ref{fig:T} and \ref{fig:A} for, respectively, the electron $n_e$ and hydrogen density $n_H$, the temperature $T$, and the metal abundance $Z$. 

\begin{figure}
    \centering
    \includegraphics[width=3.5in]{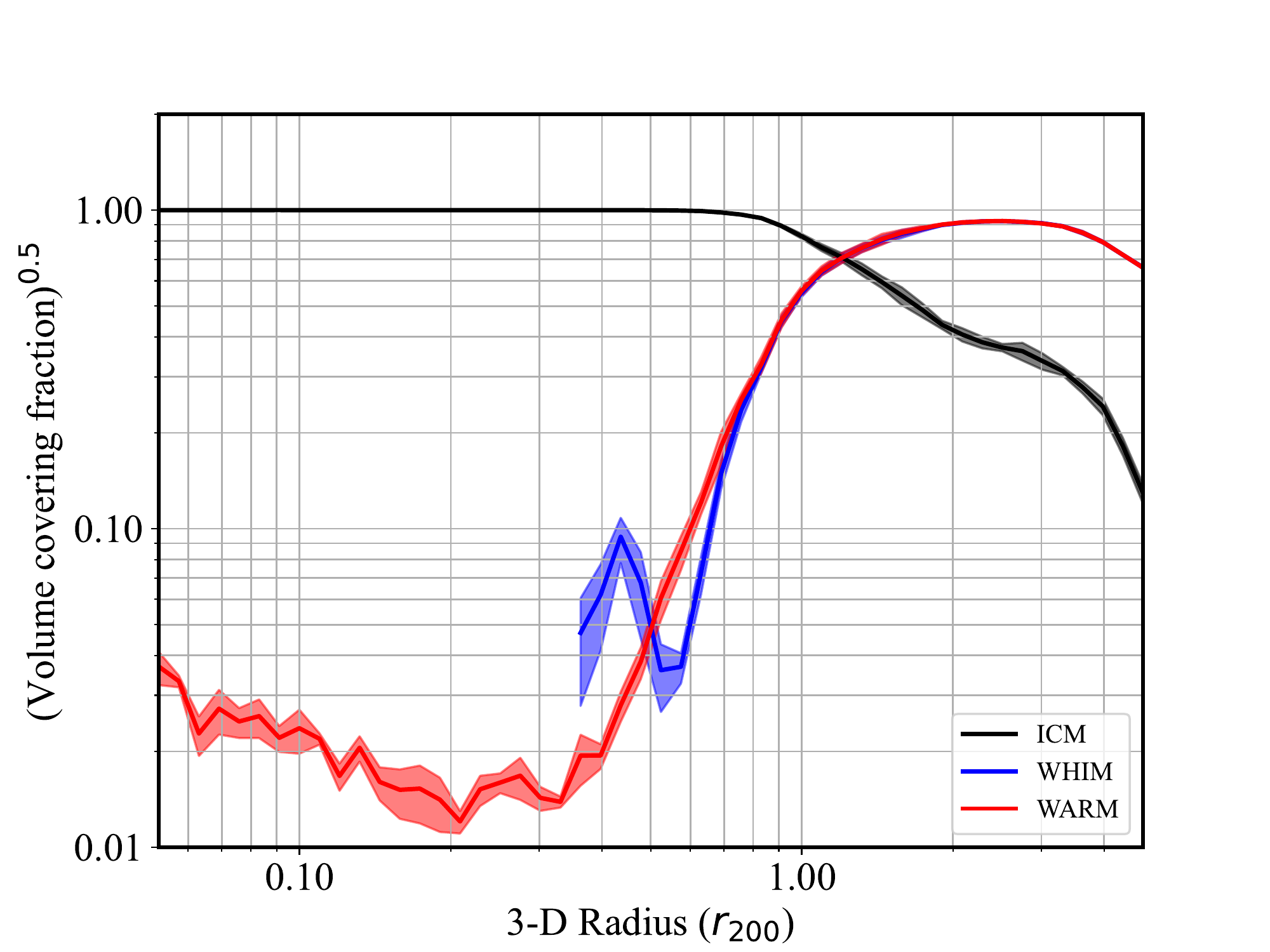}
    \caption{Volume covering fractions, same sample as in Fig.~\ref{fig:ElDens}.}
    \label{fig:V}
\end{figure}

The average profiles according to Eq.~\eqref{eq:Qaverage}, however, do not take into account the \textit{volume covering fraction} of a given phase (hot ICM, WHIM and WARM), defined as
\begin{equation}
  f_k = \dfrac{\sum\limits_{i \in R_k} V_i}{ V_{\text{shell,k}}},
  \label{eq:f}
\end{equation}
where $ V_{\text{shell,k}}$ is the usual volume of a spherical shell between radii $r_{k-1}$ and $r_k$. 
The volume covering fractions are shown in Fig.~\ref{fig:V}, where the covering fraction is approximately 100\% for the HOT phase near the cluster center, while it is significantly less than unity for the other phases, and it changes with radius for all phases. 
Indeed, the volume covering fraction of a given gas phase is intrinsically related to its abundance.
Therefore, it follows the same radial trend than the relative abundance of different gas phases discussed in \cite{gouin2022}, with two dominant phases in circum--cluster environments : the HOT gas phase (up to $\sim 1\, R_{200}$) and the WARM gas phase at larger radial distances. 

In order to predict the X-ray emission of different gas phases, it is also useful to determine the radial profiles of the mass density, accounting for the volume covered by a given phase. 
This is obtained via
\begin{equation}
  \rho_k = \dfrac{\sum\limits_{i \in R_k} m_i}{ V_{\text{shell,k}}},
  \label{eq:rho}
\end{equation}
where $m_i$ is the mass of contained in the $i$--th gas cell that is included in the radial bin $R_k$.
In Eq.~\eqref{eq:rho}, the density is in units of $(M_{\odot}/h)\cdot(\text{Mpc}/h)^{-3}$,  
and it includes all the gas (\textit{i.e.} electrons, hydrogen and all other ions). 
These mass density profiles are shown as dashed lines in Fig.~\ref{fig:ElDens}, after a normalization that results in the conversion to units of equivalent $H$ atoms cm$^{-3}$, to match the units of the electron and hydrogen density in the same figure. 
One can first notice that, when taking into account the volume fraction of a given gas phase, the radial profile of the phase drastically changes.
For example, we see that the WARM gas is denser than the ICM inside clusters (the solid red line is higher than the black one), which can be explained by the fact that colder gas can more easily collapse and become denser.
However, given that WARM gas is extremely rare compared to the ICM gas phase inside clusters (see in Fig.~\ref{fig:V} and as illustrated in Fig.~\ref{fig:Illustration_gas_dist}), its mass density (weighted by volume fraction, dashed lines) is much lower than the ICM. For a given phase, the difference between solid and dashed density curves in Fig.~\ref{fig:ElDens} is therefore simply explained by the different
radial dependencies of the volume covering fractions.

All of these thermodynamic profiles have been computed for each simulated cluster, and then averaged for all the clusters, and for the three different cluster mass bins. The detail on the averaging procedure over the cluster population has been detailed in appendix \ref{Appendix:Stat} where we also discuss the difference between mean and median averages. Indeed, individual clumps in some circum--cluster environments can artificially boost the mean thermodynamic profiles. In appendix \ref{Appendix:Tables}, we have also reported in the tables~\ref{tab:gasProperties} and \ref{tab:gasProperties2}, the values of density, volume fraction, temperature, metal abundance of the different gas phases in and around the clusters averaged over all, $M_1$, $M_2$ and $M_3$ clusters.

\subsection{Radial thermodynamic profiles - Comparisons to observations}

\begin{figure*}
    \centering
    \includegraphics[width=3.6in]{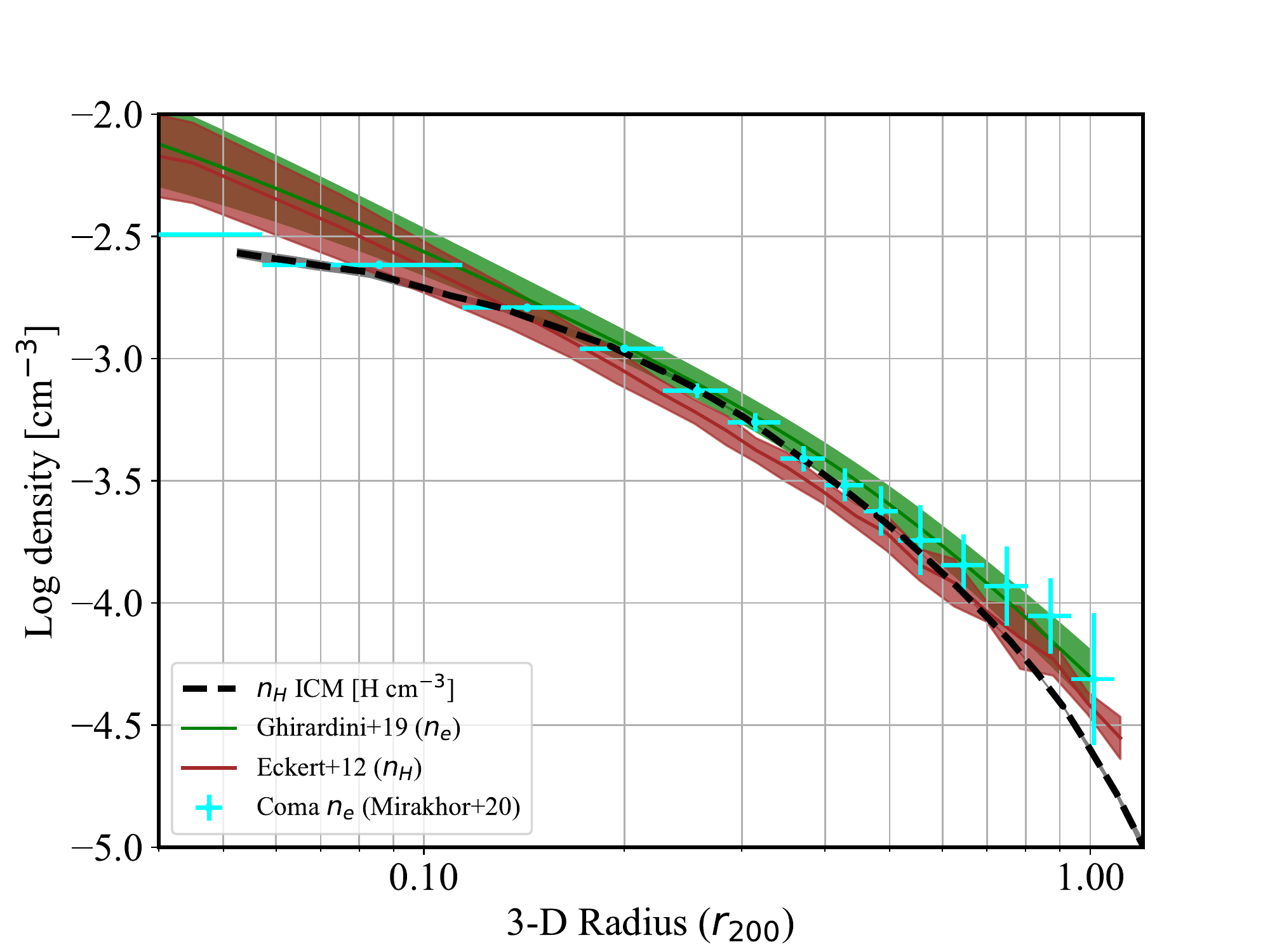}
    \includegraphics[width=3.6in]{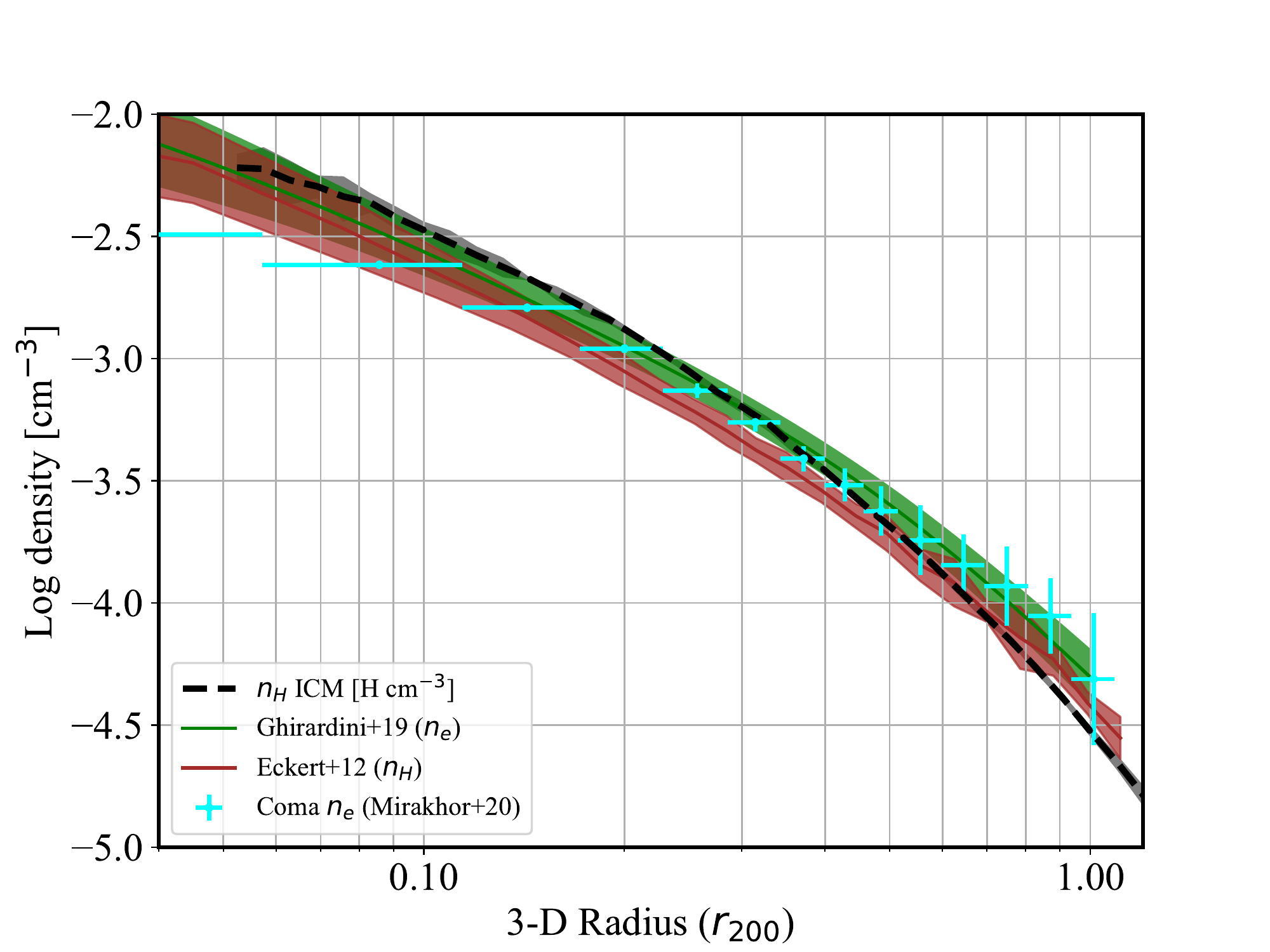}
    \caption{Zoom-in of Fig.~\ref{fig:ElDens}, the median radial profiles of the mass density for the ICM gas phase for all the clusters (left panel), and the most massive ones ($M_3$ mass bin, right panel). We highlight the comparison with X-COP observations \citep{eckert2012,ghirardini2019} and Coma cluster \citep{mirakhor2020}.}
    \label{fig:ElDensZoom}
\end{figure*}
\begin{figure*}
    \centering
    \includegraphics[width=3.6in]{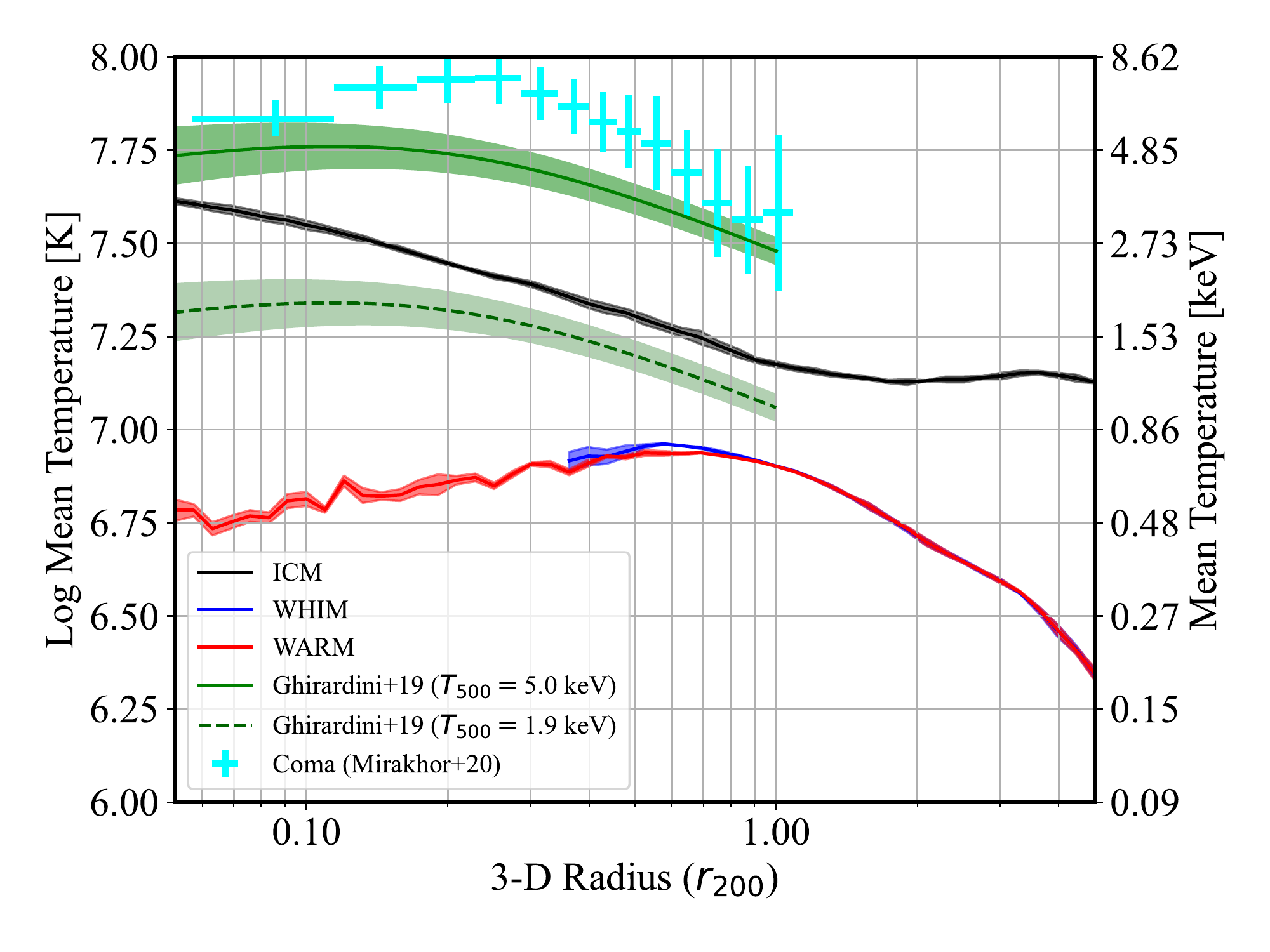}
    \includegraphics[width=3.6in]{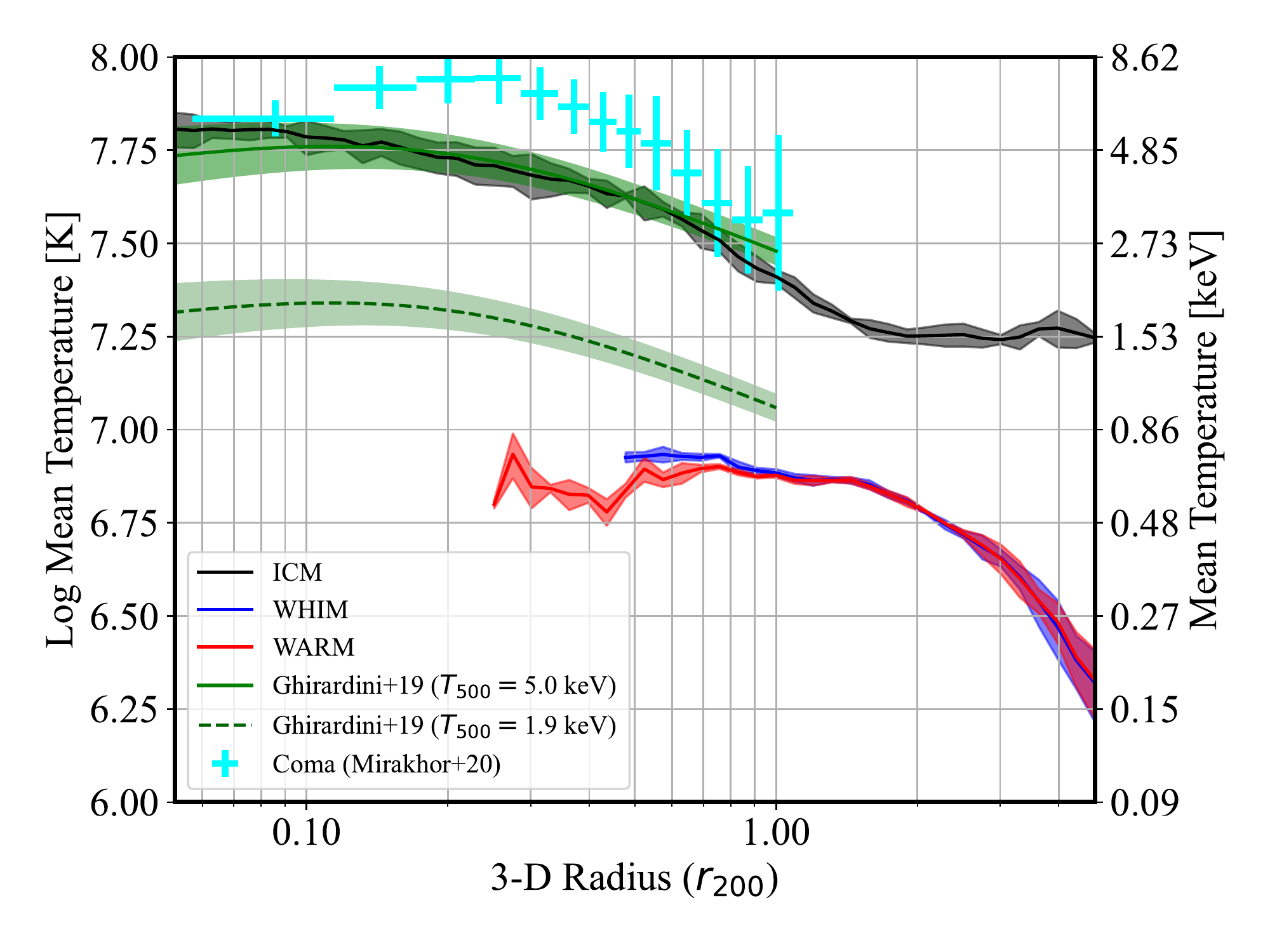}
    \caption{Radial profiles of temperatures, same samples as in Fig.~\ref{fig:ElDensZoom}.}
    \label{fig:T}
\end{figure*}
\begin{figure*}
    \centering
    \includegraphics[width=3.6in]{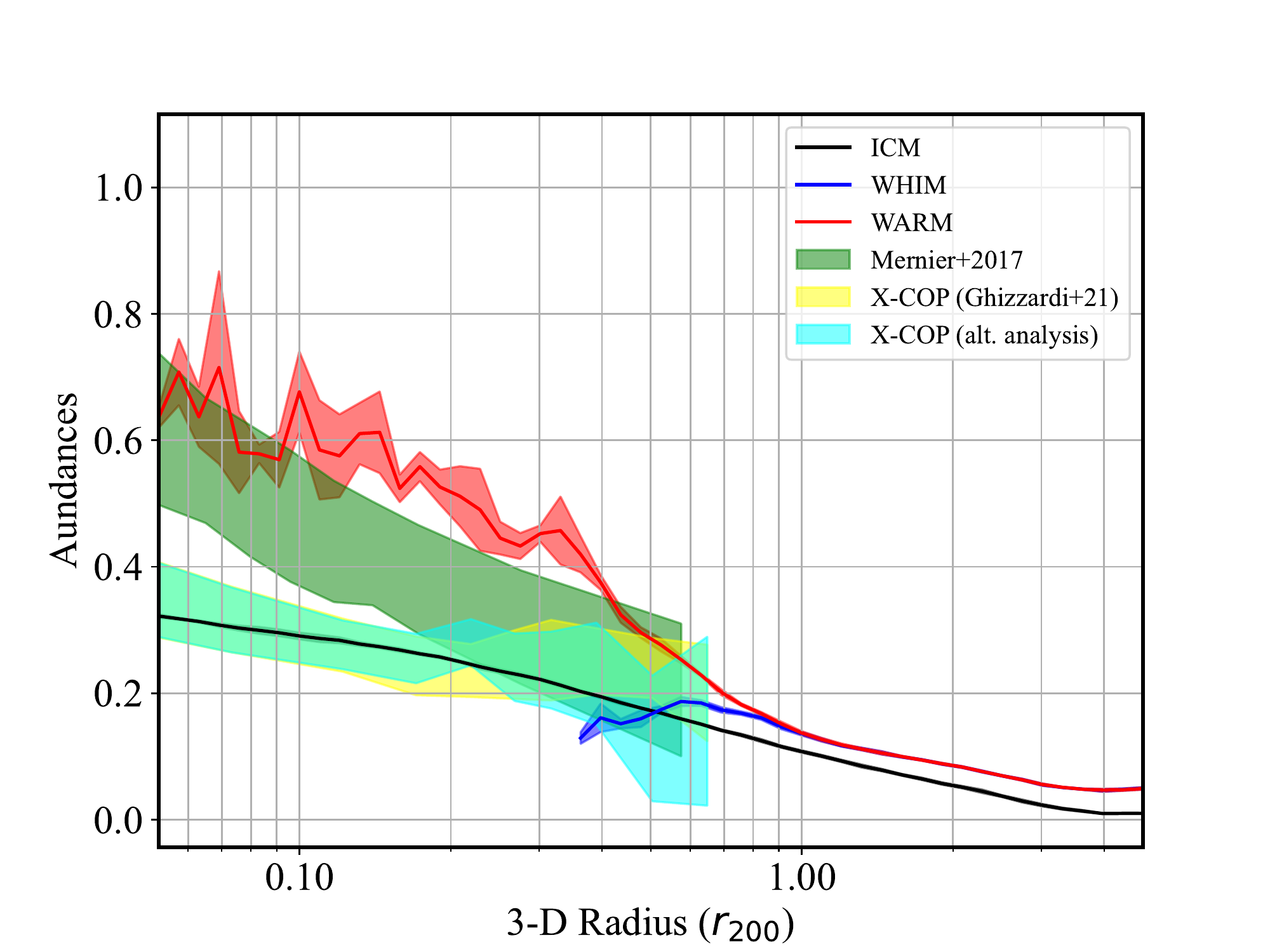}
\includegraphics[width=3.6in]{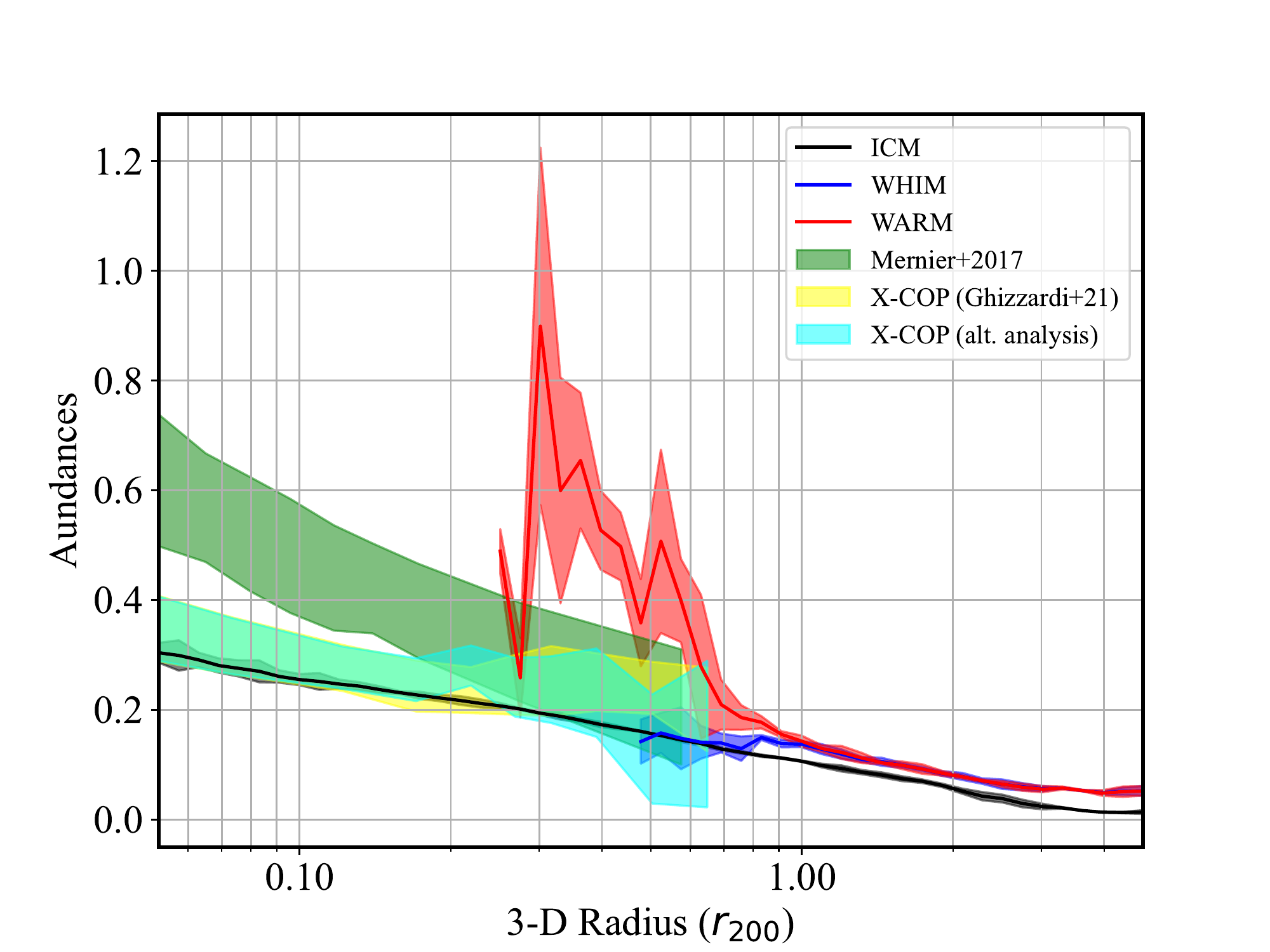}
\caption{Radial profiles of abundances, same samples as in Fig.~\ref{fig:ElDensZoom}.}
\label{fig:A}
\end{figure*}

The hot ICM mass density is compared with selected recent observations in Fig.~\ref{fig:ElDensZoom}, namely the \cite{ghirardini2019} analysis of the X-COP sample of 12 massive clusters, the \cite{eckert2012} results of a sample of 31 nearby clusters, and the detailed analysis of the Coma cluster by \cite{mirakhor2020}, all covering a radial range approximately equal to at least $r_{200}$.
The \tng\ profile of the average ICM $n_H$ density (dashed line in Fig.~\ref{fig:ElDensZoom}) is in excellent agreement with the observational results, especially for the most massive cluster sample $M_3$ whose masses are most directly comparable to those of the \cite{ghirardini2019}, \cite{eckert2012} and \cite{mirakhor2020} observations.

The radial profiles of the mean temperature are shown in Fig.~\ref{fig:T}, along with a comparison with the observational results from \cite{ghirardini2019} and \cite{mirakhor2020}. The \cite{ghirardini2019} temperature profiles are scaled according to the temperature at $r_{500}$ ($T_{500}$), and therefore Fig.~\ref{fig:T} shows two observational constraints corresponding to two values of $T_{500}$ that correspond approximately to the most massive sample and to the least massive sample. 
The radial profile of the hot ICM temperature for the $M_3$ sample is in perfect agreement with the \cite{ghirardini2019} constraints for $T_{500}=5$~keV. The Coma radial profile has a slightly higher temperature, given that its mass of $M_{200}=8.550\pm0.55$~M$_{\odot}$ corresponds to one of the largest in our \tng\ sample.

The average metal abundance profiles are shown in Fig.~\ref{fig:A}. The metal abundance of the ICM is in excellent agreement with the \cite{ghizzardi2021} constraints at all radii. \cite{mernier2017}, on the other hand, finds a significantly higher metal abundance at lower radii, progressively reaching the $A \simeq 0.2$~Solar value towards the virial radius, in agreement with the \tng\ profiles. This higher metallicity inside clusters might be induced by WARM gas phase, as it is suggested by our predictions according to Fig.~\ref{fig:A}. 

\subsection{Clumpiness of the gas}
\label{sec:clumpiness}

Inhomogenieties in the gas have a direct impact on several physical quantities derived from observations of galaxy clusters and the warm--hot phases. As is explained in detail in Sect.~\ref{sec:SX} below, the measured
X--ray surface brightness is proportional to the square of the density. 
Therefore, it is the average $\langle n_e^2  \rangle$, and not $\langle n_e \rangle$, that is the primary observable in X--rays. In turn, this means that inferences on such fundamental quantities as the gas mass and the total mass as inferred by hydrostatic equilibrium, both function of $n_e$,
rely on an estimate of the linear density and therefore on an assessment of inhomogenieties in the plasma \citep[for a review, see e.g.][]{walker2019}. 

Following \cite{nagai2011}, we use the following averages of the electron density, 
\begin{equation}
\begin{cases}
     \langle n_e  (r_{k})  \rangle = \dfrac{\sum\limits_{i \in R_k} {n_e}_i \times V_i}
     {\sum\limits_{i \in R_k} V_i}\\[15pt]
     \langle {n_e}^2  (r_{k}) \rangle = \dfrac{\sum\limits_{i \in R_k} {{n_e}_i}^2  \times V_i}{\sum\limits_{i \in R_k} V_i}
     \end{cases}
\end{equation}
to define the \textit{clumping factor} as
\begin{equation}
    C (r_{k}) = \dfrac{   \langle {n_e}^2  (r_{k})  \rangle  }{ \langle n_e  (r_{k})  \rangle^2 }\geq 1\,.
    \label{eq:C}
\end{equation}

A value of the clumping factor greater than one indicates gas that is not uniformly distributed over the respective spherical shell of provenance, but rather with regions of significantly higher or lower (including null) density, contributing to a variance in the distribution of the densities that results in $ \langle {n_e}^2  (r_{k})  \rangle > \langle n_e  (r_{k})  \rangle^2$. In Appendix \ref{Appendix:Tables}, the table~\ref{tab:gasProperties3} also reports the values of $\sqrt{C}$ for selected radii, for the three gas phases and for the 3 mass bins. 

\begin{figure}
    \centering
    \includegraphics[width=3.5in]{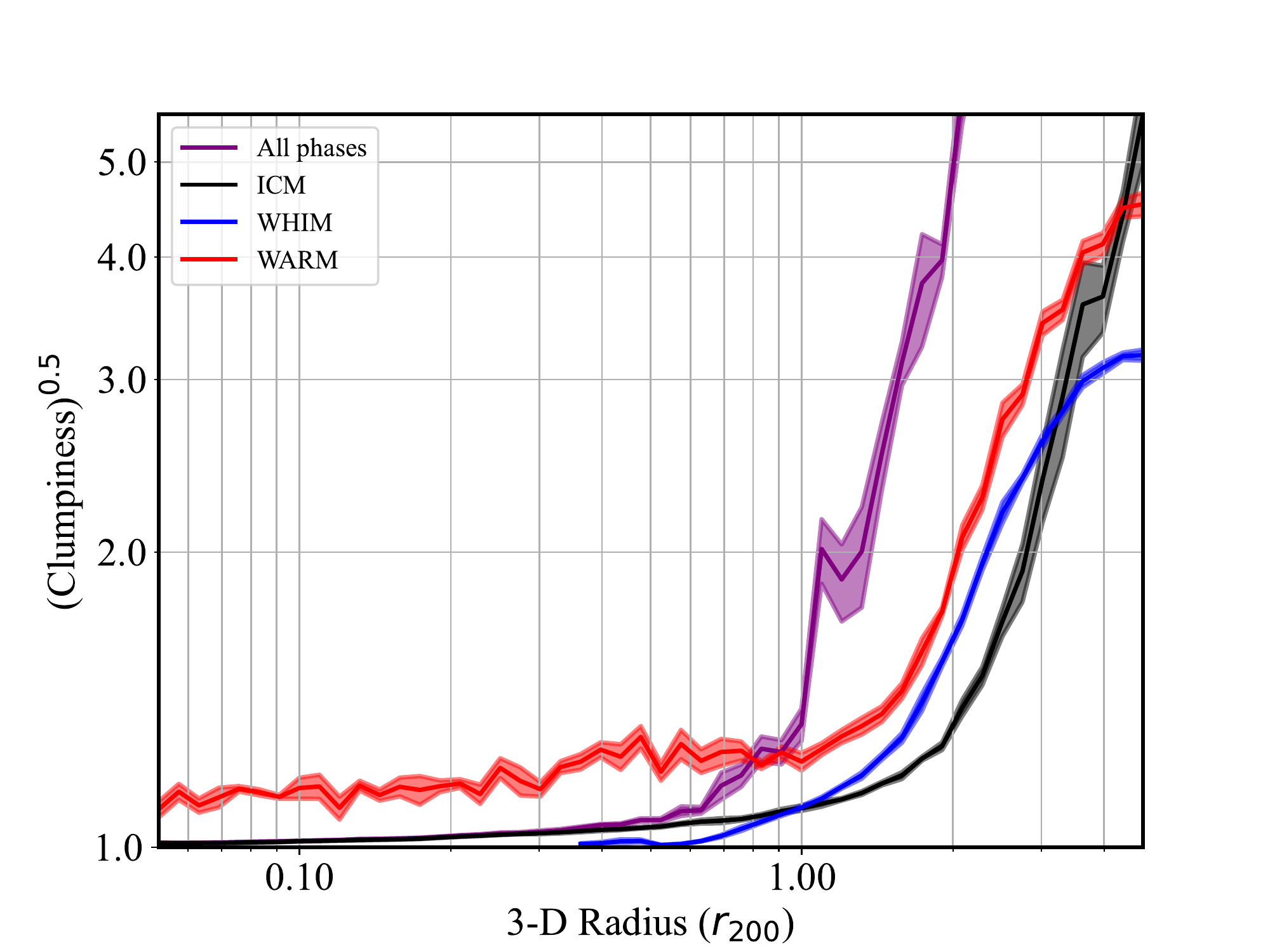}
\caption{Median radial profiles of the clumping factor average over all the clusters. }
    \label{fig:clumping}
\end{figure}

Figure \ref{fig:clumping} shows the median radial profile of the clumping factor average over all the clusters, and for different gas phases. Considering the full gas distribution (purple line), the clumpiness of the gas is very low inside clusters and becomes significantly clumpy beyond the virial radius, while the density of the hot ICM is quite uniform within $r_{200}$. However, the WARM gas clumpiness (red line) is higher inside clusters, as suggested previously in the illustration Fig.~\ref{fig:distr}. Therefore, we conclude that the WARM gas phase is colder,
with larger metallicity and more clumpy than the ICM in circum--cluster environments. In contrast, WHIM gas only is by definition more diffuse, and thus with a lower clumping factor than WARM gas. Interestingly, we found also that all gas phases follow the same radial trend with low clumpiness inside clusters, and rapidly increasing beyond the virial radius.

The measurements of the clumping factor of \cite{nagai2011}, using the simulations of \cite{kravtsov1999,kravtsov2002}, shows an increasing radial trend that reaches a value of $ C \simeq 10$ at $2 \times R_{200}$, somewhat higher than our results yet qualitatively consistent with an increasing importance of clumpiness at larger distances from the cluster's center. This radial evolution of clumpiness is also consistent with previous numerical investigations \citep[see e.g.][]{Roncarelli2013,Ansarifard2020}. More recently, \cite{angelinelli2021} has performed a similar analysis with the \texttt{Itasca} simulations (see Fig.~5 in their paper), resulting in quantitatively similar results that indicate an increasing radial trend of the clumpiness with radius reaching an approximate value of $\sqrt C\simeq 3$ at $5 \times r_{500} \simeq 3 \times r_{200}$ (their definition of the clumpiness factor includes the square--root operation). An interesting point is to consider either mean or median averaging of clumping factor, a point discussed in Appendix \ref{Appendix:Stat}, showing that median is less sensitive to individual gas inhomogeneities in the full cluster sample \citep[see also ][reaching a similar conclusion]{Towler2023}. Finally, \cite{Planelles2017} have revealed that the inclusion of AGN feedback in simulations is helping the ICM to be more homogeneous, reducing the clumping factor values, and found similar clumpiness radial trend, with a slightly higher value of $\sqrt C = 1.2$ at $r_{200}$.

Observationally, it is challenging to measure the gas clumping directly,  primarily due to the limited spatial resolution of the X--ray instruments and the small angular size of most distant clusters.
\cite{morandi2017} was able to estimate a clumping factor of $C \sim 2-3$ at $r_{200}$ for the lower--mass galaxy group NGC~2563, again in agreement with the \tng\ predictions, although for a system of lower mass than the clusters
in our sample. For the nearby Virgo cluster, \cite{mirakhor2021} measured a small value of the clumping  factor ($C \simeq 1.1$) at small radii from the center ($r \ll r_{200}$) from X--ray observations, again in quantitative agreement with our measurements.
Moreover, \cite{Eckert2015b} used a sample of 31 clusters ($z=0.04=-0.2$) observed by ROSAT to measure the clumping factor at the level of $\sqrt{C} \leq 1.1$ within $r_{500}$, and to set constraints of the order $\sqrt{C} \leq 1.5-2$ beyond that radius.

\begin{figure}
    \centering
    \includegraphics[width=3.5in]{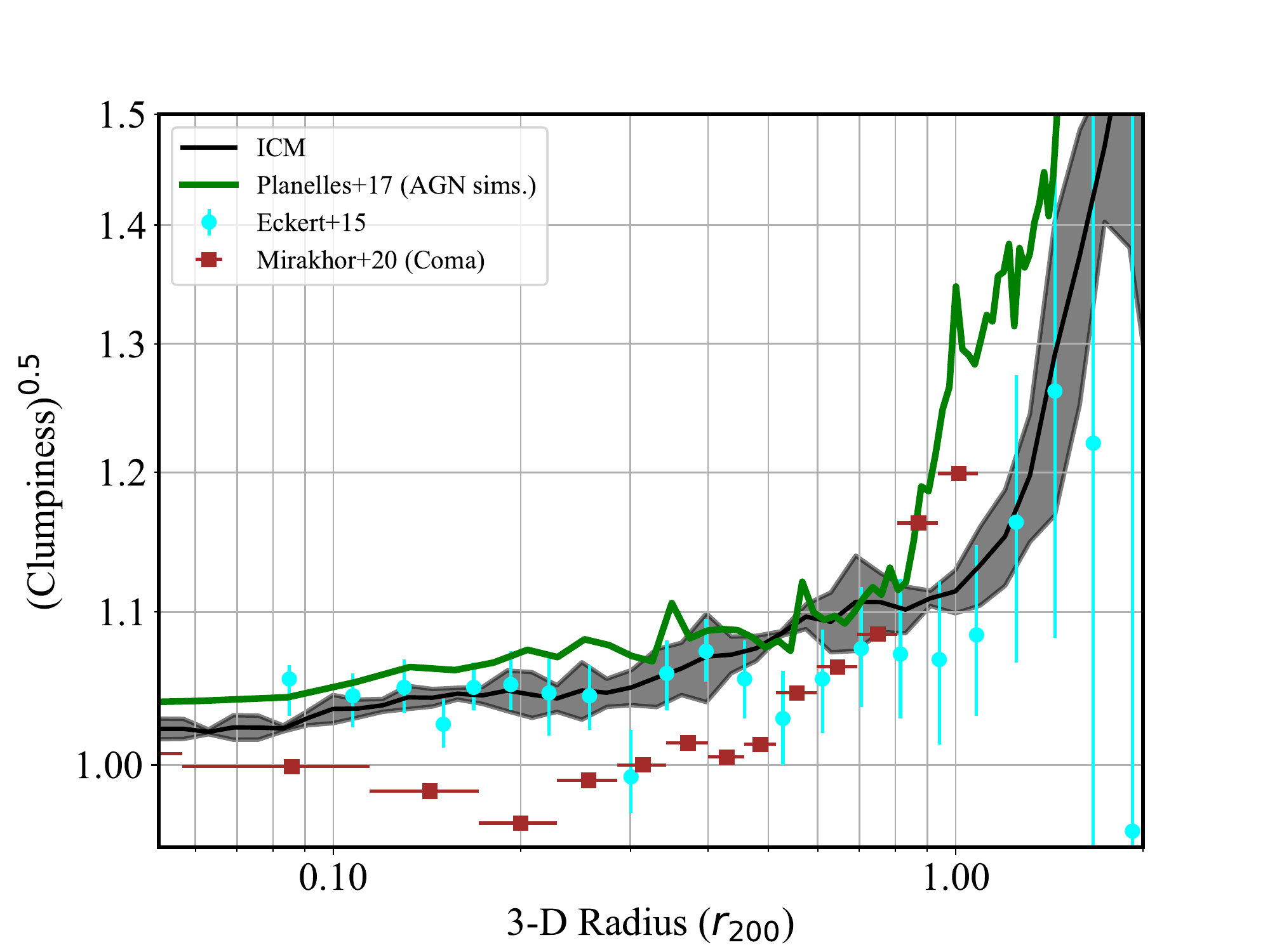}
    \caption{Comparison of the ICM clumpiness factor for the most massive cluster ($M_3$ mass bin) with the simulations of \cite{Planelles2017} in a sample of clusters in a comparable mass range, the clumpiness in Coma from \cite{mirakhor2020} and the \cite{eckert2015} measurements.}
    \label{fig:clumping-zoom}
\end{figure}

In Fig.~\ref{fig:clumping-zoom}, we compare our ICM clumpiness radial profile of most massive clusters with previous AGN simulation from \cite{Planelles2017}, and the observational measurements from \cite{Eckert2015b} and \cite{mirakhor2021}. Inside clusters ($<1\,R_{200}$), a good agreement is found with both previous numerical predictions and recent observations. Interestingly, at larger radii ($>1\, R_{200}$),  \cite{Planelles2017} show somewhat lower clumping factor than our prediction, although with the same increasing 
radial trend as our profile. This difference can be due to different AGN modeling, which accurately reproduces observed statistical trends at $z=0$ in \tng\ \citep{Pillepich2018,Nelson2019}. The accurate AGN feedback modeling in \tng\ might be 
the reason for the  agreement of our prediction of the ICM clumpiness with recent observations beyond the virial radius.

\subsection{The gas mass fraction}

Finally, we report a first analysis of the gas mass fraction our cluster sample, to explore its validity in terms of the baryon mass budget, and compare it to current observational measurements. 
The baryonic budget of \tng\ simulations \citep{Nelson2019} follows the cosmological parameters from the Planck 2015 results \citep{planck2016-cosmology}. 

\begin{figure}
    \centering
    \includegraphics[width=0.5\textwidth]{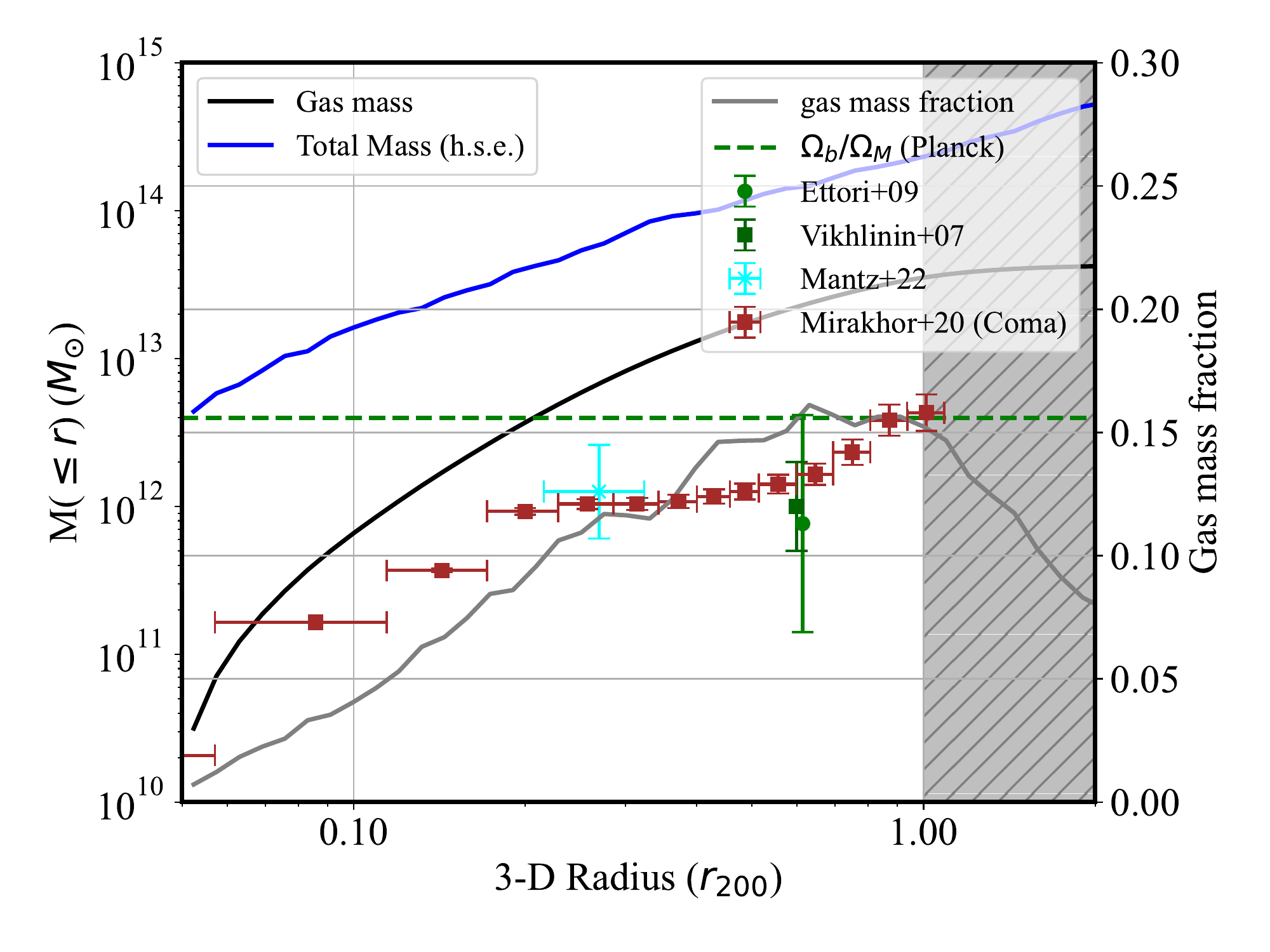}
    \caption{Cumulative radial profiles of gas mass and hydrostatic mass for the entire cluster sample. The X--ray measurements of \cite{ettori2009} and \cite{vikhlinin2006} are for $r\leq r_{500}$, and the \cite{mantz2022} gas mass fraction was calculated in a radial shell $0.8-1.2 \times r_{2500}$ , where $r_{2500}\simeq 0.45 \times r_{500}$.}
    \label{fig:fgas}
\end{figure}

In our case, we use the average profiles of the ICM density and temperature to estimate the gas mass fraction, under the assumption of hydrostatic equilibrium. Gas mass fractions can be used to probe the hydrostatic mass bias in observations \citep{Wicker2022}. Therefore, we focus here only on the gas cells, and estimate the total mass according to the hydrostatic equilibrium \citep[e.g.][]{sarazin1988} within a radius $r$ as
\begin{equation}
    M(r) = - \dfrac{K T(r) r}{\mu m_p G} \left( \dfrac{d \ln n_e}{d \ln r} + \dfrac{ d \ln T}{d \ln r} \right),
\end{equation}
where $\mu$ is the mean molecular weight of the plasma (see Sect.~\ref{sec:emission} for details), and
$n_e$ and $T$ are the electron density and temperature at a give radius $r$ from the center of the clusters.
Figure~\ref{fig:fgas} shows the radial profile of the gas mass and of the total mass, for the entire cluster sample.
Within the virial radius, the gas mass fraction is in excellent agreement with the ratio of the cosmological parameters $\Omega_b/\Omega_M$ as measured by the Planck mission \citep{planck2020}, and from various X--ray measurements of the gas mass fraction \citep[e.g.][]{mantz2022,mantz2014,ettori2009,vikhlinin2006,Lyskova2023}. 
This agreement is further evidence that the definition of the ICM phase in the \tng\ simulation used in this work does indeed reproduce key features of observed clusters.

For completeness, Fig.~\ref{fig:fgas} includes the radial range beyond $r_{200}$ where it is known that hydrostatic equilibrium is challenged by the presence of non--thermal pressure support \citep[e.g.][]{Fusco-Femiano2018}. 
The application of hydrostatic equilibrium to regions where gas is supported by additional forms of pressure results in a higher--than--true total mass, and in fact our radial profile of the gas mass fraction has a sharp drop. 
In principle, our analysis can be complemented by a study of the radial distribution of dark matter in \tng, which, however, goes beyond the scope of this paper \citep[see][for details on hydrostatic mass bias for mock X-ray in IllustrisTNG simulation]{Barnes2021}.


\section{Projection of the radial profiles}
\label{sec:projection}
The radial profiles of thermodynamical properties in Fig. \ref{fig:ElDens} up to  Fig. \ref{fig:clumping} are a function of the three--dimensional radius, as measured by the \tng\ simulation. 
Observations of the thermodynamic quantities, such as density and temperature, and of the associated X--ray surface brightness (see Sect. \ref{sec:SX}) require either a \textit{deprojection} of the observations (as, e.g., is done in the case of observations
such as those of \citealt{ghirardini2019}), or a \textit{projection} onto a (focal) plane of the three--dimensional profiles obtained from the simulations.  This section described the method
of projection that we use for this project, so that we can compare our results with
X--ray observations directly, without the need to deproject the observations.

The projection of the radial profiles follows the geometry of Fig.~\ref{fig:onion}, where the 
spherically--symmetric distribution of matter is represented by onion--like shells of increasing
radial distance $r$ from the center. The usual spherical coordinates $(r,\theta,\phi)$ are related to the 
cylindrical coordinates $(\rho,\theta,z)$ by 
\begin{equation}
  \begin{cases}
    \rho=r \sin \phi \\
    z = r \cos \phi,
  \end{cases}
\end{equation}
where the $z$ direction in cylindrical coordinates represents the sightline.
Given the spherical symmetry of the radial profiles, the $z$ coordinate of the 
cylindrical system is chosen to represent the distance along the sightline, and the $(\rho,\theta)$ plane represents the focal plane.
As illustrated in Fig.~\ref{fig:onion}, the volume of intersection of a spherical shell between radii $\RminEq=r_{k-1}$ and $\RmaxEq=r_k$ and a concentric cylindrical shell between radii $\rhominEq$ and $\rhomaxEq$ can be calculated using the following steps. 

\subsection{The volume of intersection of a sphere with a concentric cylinder}

This initial problem consists of finding the volume of intersection of a whole sphere of radius $R$ with a concentric cylinder of radius \rhomax. If $\rhomaxEq \geq R$, the intersection if the entire sphere, so the interesting case is when $\rhomaxEq \leq R$.

The volume element in cylindrical coordinates is $dV = r\; dr\; d\theta dz$, and the limits of integration
to obtain the volume of intersection are
\begin{equation}
  \begin{cases}
    \rho \in [0, \rhomaxEq]\\
    \theta \in [0, 2 \pi]\\
    z \in [0, z(\rho,R)]\\ 
  \end{cases}
\end{equation}
where $z(\rho,R)$ represents the maximum value of the Cartesian $z$ coordinate for a fixed value of the
$\rho$ cylindrical radius, and it is given by
\begin{equation}
z(\rho,R)^2 + \rho^2 = R^2.
  \label{eq:zmax}
\end{equation}
Eq.~\ref{eq:zmax} means that, as the cylindrical coordinate spans its range between 0 and \rhomax,
the maximum value of $z$ for integration decreases
from the sphere's radius $R$ at the center ($\rho=0$), to a smaller value $z(\rhomaxEq,R) = \sqrt{R^2 - \rhomaxEq^2}$,
of course in the interesting case of $\rhomaxEq \leq R$.

This leads to a volume of intersection of
\begin{equation}
  V(\rhomaxEq,R) = 2 \int \limits_0^{2 \pi} d \theta \int \limits _0^{\rhomaxEq} \rho d\rho \int \limits_0^{\sqrt{R^2 - \rho^2}}  dz
\end{equation}
which can be immediately solved with a change of variable $u=R^2-\rho^2$, $du=-2 \rho d\rho$, again assuming $\rhomaxEq \leq R$, whereby
\begin{equation}
\begin{aligned}
  V(\rhomaxEq,R) =& 4 \pi \int \limits_{R^2}^{R^2-\rhomaxEq^2} \left( - \sqrt{u}\, \dfrac{du}{2} \right) = \\
  &  - \left. \dfrac{4}{3}  \pi u^{\nicefrac{3}{2}} \right|_{R^2}^{R^2-\rhomaxEq^2}.
  \end{aligned}
\end{equation}

\begin{figure}
  \centering
  \includegraphics[width=3.5in]{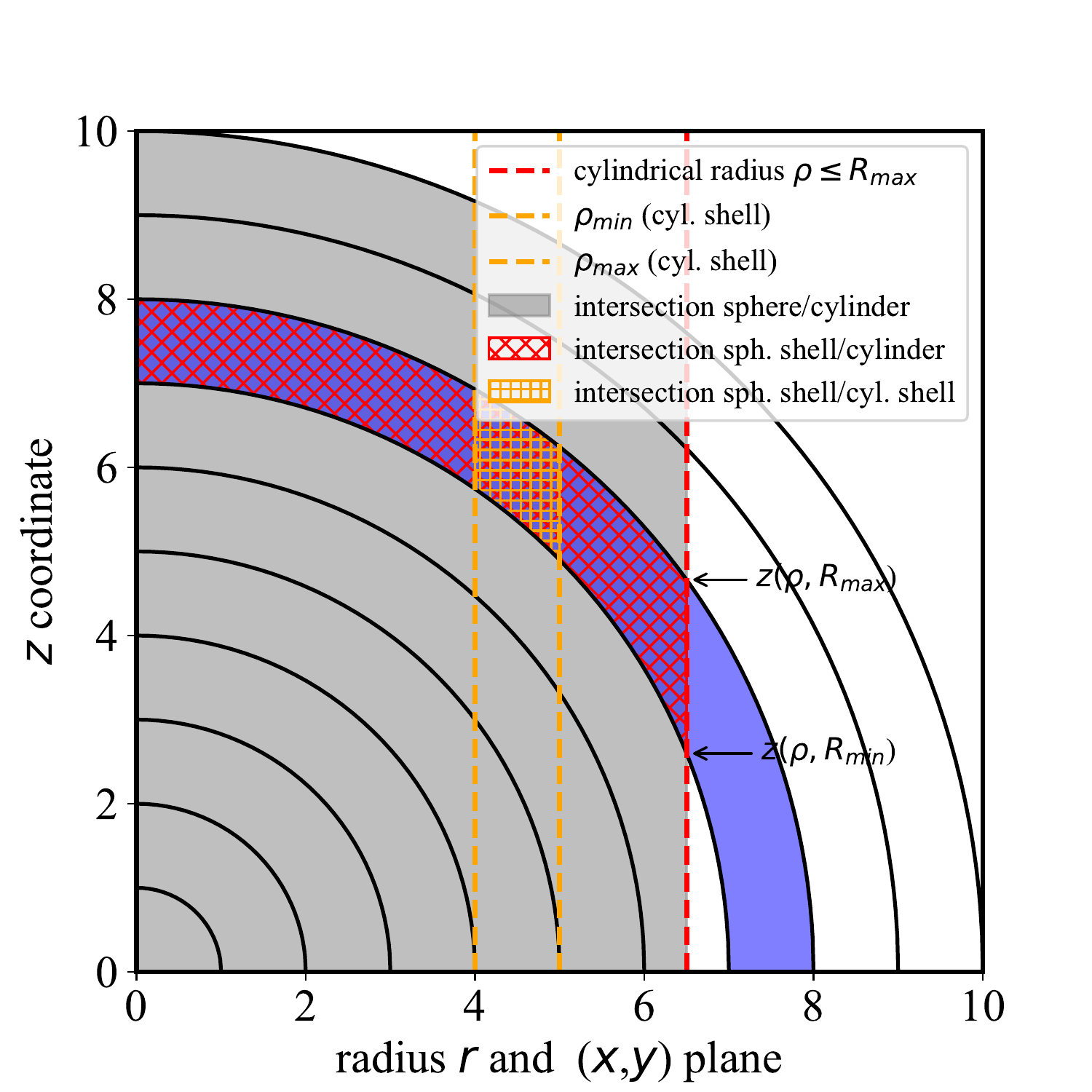}
  \caption{Volumes of intersection for the projection of
a three--dimensional radial distribution onto a two--dimensional plane.
  In grey is the volume of intersection of concentric sphere and cylinder.
  In blue is the volume of intersection of a spherical shell with radii \Rmin, \Rmax\
  with a cylinder of
  radius \rhomax (in red). The volume of interaction represents the volume contribution from the spherical shell (assumed to
  have uniform thermodynamical properties) to
  a circular region of radius $\rho$ (dashed red line). For a cylindrical shell of radii \rhomin\ and \rhomax\ (in orange),
  the area of intersection represents the contribution to the annulus by the shell. Given the azymuthal symmetry of both sphere
  and cylinder, the polar angle $\theta$ is not displayed.}
  \label{fig:onion}
\end{figure}

This integral leads to the results that the volume of intersection of a sphere of radius $R$ with a concentric
cylinder of radius \rhomax\ is given by
\begin{equation}
\begin{aligned}
  V(\rhomaxEq,R) =&\\  \dfrac{4}{3} \pi \times &
  \begin{cases}
    \begin{aligned}
      & R^3& \text{ if $R \leq \rhomaxEq$}\\[5pt]
      & \left(R^3 - (R^2-\rhomaxEq^2)^{\nicefrac{3}{2}} \right) & \text{ if $R \geq \rhomaxEq$}.
    \end{aligned}
 \end{cases}
 \end{aligned}
  \label{eq:VrhoR}
\end{equation}
This volume is shown as the grey area in Figure~\ref{fig:onion}.

\subsection{Volume of intersection of a spherical shell with a concentric cylinder}

The next problem is to intersect a spherical shell with radii \Rmin, \Rmax, with the same concentric cylinder of
radius \rhomax. The volume of interaction represents the volume contribution from the spherical shell to
a circular region of radius \rhomax\ in the plan of the detector.

For this volume, the cylindrical coordinates have the following ranges:
\begin{equation}
  \begin{cases}
    \rho \in [0, \rhomaxEq] \text{ (same as in previous case)}\\
    \theta \in [0, 2 \pi] \text{ (same as in previous case)}\\
    z \in [z(\rho,\RminEq), z(\rho,\RmaxEq)]
    \label{eq:limits}
  \end{cases}
\end{equation}
where the new limits  for the $z$ coordinate enforce the two radii of the spherical shell. This geometry is
represented in Figure~\ref{fig:onion}.

With the limits of integration according to Eq. \ref{eq:limits}, the volume of intersection is given by
\begin{equation}
  V(\rhomaxEq,\RminEq,\RmaxEq)= 4 \pi \int\limits_0^{\rhomaxEq} \rho d\rho  \int\limits_{\sqrt{\RminEq^2-\rho^2}}^{\sqrt{\RmaxEq^2-\rho^2}} dz
  \label{eq:VrhoRmaxRmin1}
\end{equation}
provided that $\rhomaxEq \leq \RminEq$, so that the lower limit of integration remains described by Eq. \ref{eq:zmax}.
The case of $\RminEq \leq \rhomaxEq \leq \RmaxEq$ (i.e., in Figure~\ref{fig:onion} the red dashed line would be
between the values $r=7$ and 8) is not used in practice, and it will be discussed separately.
This integral is given by

\begin{equation}
\begin{aligned}
  V&(\rhomaxEq,\RminEq,\RmaxEq)=\\[10pt] & 
  \begin{cases}
    \begin{aligned}
      & \dfrac{4}{3} \pi \left(\RmaxEq^3 - \RminEq^3\right) & \text{ if $\rhomaxEq \geq \RmaxEq$}\\[5pt]
      & V(\rhomaxEq,\RmaxEq)-V(\rhomaxEq,\RminEq) & \text{ if $\rhomaxEq \leq \RminEq$}\\[5pt]
     & \text{(unused)} \text{ if $\RminEq \leq \rhomaxEq \leq \RmaxEq$} & \\
    \end{aligned}
  \end{cases}
  \end{aligned}
  \label{eq:VrhoRmaxRmin2}
\end{equation}

In Eq. \ref{eq:VrhoRmaxRmin2}, the top case corresponds to a spherical shell that is completely enclosed by the
cylinder. The middle case corresponds to the situation illustrated in Figure~\ref{fig:onion}, where
Eq. \ref{eq:VrhoRmaxRmin1} applies, and therefore the integral is the difference of the two contributions
given by Eq. \ref{eq:VrhoR} (which has the limitation $\rhomaxEq \leq \RminEq$). The last case is not used,
since we chose the boundaries of the annuli to be the same as those of the spherical shells.

\subsection{Volume of intersection of a spherical shell with a concentric cylindrical shell}

Finally, the volume of interest is the intersection of a spherical shell with a concentric
cylindrical shell of radii \rhomin, \rhomax, and it is
represented by the orange shaded area in Figure~\ref{fig:onion}. To simplify calculations, we chose the same
radial bins for the three--dimensional radius $r$ and for the projected cylindrical radius $\rho$, in such a way that
the radii \rhomin, \rhomax\ always fall at the boundaries of the \Rmin, \Rmax\ spherical radii (i.e., as
for the orange dashed lines in Figure~\ref{fig:onion}, but not the red line, which was used only for
illustration purposes). With this choice, it is also not necessary to evaluate the case at the bottom of Eq. \ref{eq:VrhoRmaxRmin2}.

Given the linearity of the integral, the volume of intersection is given by
\begin{equation}
  \begin{aligned}
     V&(\rhominEq,\rhomaxEq,\RminEq,\RmaxEq)= \\
     V&(\rhomaxEq,\RminEq,\RmaxEq)-V(\rhominEq,\RminEq,\RmaxEq)
  \end{aligned}
  \label{eq:VrhoMaxrhoMinRmaxRmin}
\end{equation}
as the difference of the two contributions according to Eq. \ref{eq:VrhoRmaxRmin2}. This volume is
used to integrate the total power or luminosity emitted by a given radial region along the sightline, given that the radiation is expected to be optically thin and therefore
the surface brightness adds linearly along the sightline. The use of these equations is explained in the following section.


\section{X--ray emission from \tng\ circum--cluster environments: Methods and preliminary results}
\label{sec:SX}
In this section we present the methods used to calculate and project 
the X--ray emissivity from the simulated clusters, and preliminary results of our analysis.

\subsection{X-ray cooling function, surface brightness and luminosity}

 In X--ray astronomy, it is customary to refer to the power $P$ emitted by an astrophysical
 plasma, in units of ergs~s$^{-1}$, as the luminosity of the source. This luminosity is
 related to the plasma \textit{emissivity} $\Lambda(T,A)$ or \textit{cooling function}, which is defined as the power emitted
 per unit $n_e n_H$ by the plasma, following the convention of \cite{raymond1977} and \cite{smith2001},
 and therefore in units of ergs~cm$^{3}$~s$^{-1}$. The plasma emissivity is therefore the power emitted
 by a unit volume of the plasma, rescaled by the $n_e n_H$, and it is a function
 of the temperature $T$ and metal composition of the plasma, the latter usually described as a fraction
 $A$ of Solar abundances of heavy elements. Both the \tng\ simulations and the \texttt{APEC} code of \cite{smith2001}  use the \cite{anders1989} abundances, and therefore those metal abundances are used in this paper.~\footnote{The X--ray emissivity is calculated using the \texttt{ATOMDB} and \texttt{APEC} databases, as implemented in the \texttt{pyatomdb} software.}

 The plasma cooling function is related to the X--ray surface brightness via 
\begin{equation}
S_X = \int_l n_e n_H \Lambda(T,A)\, dl\;\; \text{(erg~cm$^{-2}$~s$^{-1}$)}
\label{eq:Sx}
\end{equation}
where the integral is along the sightline $l$, and it is intended to represent the
amount of energy collected, per unit time, in a unit area of an ideal detector in the focal
plane of a focusing telescope. This is the quantity that is usually collected in X--ray observations of diffuse sources such as galaxy clusters \citep[e.g.][]{ghirardini2019}.
Another observational quantity commonly used is the \textit{emission integral}, sometimes referred
to as \textit{emission measure}, defined by
\begin{equation}
    I = \int_V n_e n_H\, dV\;\; \text{(cm$^{-3}$)}
    \label{eq:EI}
\end{equation}
and representing an integral over a volume $V$ of an extended source. The emission integral
is proportional to the normalization constant of an optically--thin plasma, such as \texttt{APEC}, used to fit a spectrum from a spatially--resolved area in the detector (e.g., a radial annulus around the center of the cluster), with $V$ representing the three--dimensional volume responsible for the emission. For example, the \texttt{APEC} code (and other commonly used codes such as \texttt{mekal}, \citealt{mewe1985,mewe1986,kaastra1992}) as implemented in the X--ray fitting software \texttt{XSPEC} returns a normalization constant
\begin{equation}
    \text{Norm} = \dfrac{10^{-14}}{4 \pi (D_A(1+z))^2} \times I 
\end{equation}
where $D_A$ is the angular diameter distance to the source, $z$ its redshift, and the constant $10^{-14}$ is used for convenience so that the normalization constant has values in the neighborhood of unity for a variety of astronomical sources of interest, including 
galaxy clusters.

For the purpose of comparing the amount of energy radiated by different phases, and at
different projected radial distance from the center of the cluster, it is necessary to use quantities that are closely related to observations. We assume radial
symmetry throughout this work, and therefore consider the three--dimensional
radius $r$ as the only geometric parameter, with $R_k$ a radial
range between three--dimensional radii $r_{k-1}$ and $r_k$, as used in Sect.~\ref{sec:profiles}.
The total power radiated within a \textit{projected} radial range $C_k$ (i.e., 
projected radii $\rho_{k-1}$ and $\rho_k$), is thus an integral along the sightline
representing a cylindrical shell within those projected radii, 
\begin{equation} 
    P_k = \int_{C_k} \Lambda( {\scriptstyle T(r),A(r)}  ) \cdot n_e(r)\cdot n_H(r) \cdot dV \;\; \text{(erg~s$^{-1}$)}
    \label{eq:P}
\end{equation}
where the volume $V$ is the intersection between the sphere and the concentric 
cylindrical annulus. In practice, given that three--dimensional radial profiles 
are evaluated in discrete concentric shells, the power can be evaluated as a sum over all volumes
resulting from the intersection of the cylindrical shell (representing the sightline) with the spherical shells (representing the cluster), i.e.,
\begin{equation} 
    P_k \simeq \sum\limits_{i=1}^N \Lambda_i \cdot n_{e,i}\cdot n_{H,i} \cdot V_{ik} \;\; \text{(erg~s$^{-1}$)}
    \label{eq:Pdiscrete}
\end{equation}
where $V_{ik}$ is the volume of intersection of cylindrical shell with radii in the fixed $C_k$ range
with a spherical shell with radii in the variable $R_i$ range, as described in Sect.~\ref{sec:projection}, Eq.~\ref{eq:VrhoMaxrhoMinRmaxRmin}.
This projected power is proportional to the \textit{count rate} of photons that are detected by a given instrument, and in a given band, and therefore is a useful proxy to establish the relative level of emission of the phases as a function of projected radius.

In Eq.~\ref{eq:Pdiscrete}, $V_{ik}$ is the volume of the intersection between the $i$--th spherical shell of density $n_{e,i}$,
and a cylindrical shell with inner and outer radii representing the $k$-th annulus on the focal plane, the latter were chosen for convenience as the same as those of the $k$--the spherical shell. 
Details of these volumes of intersection were given in Sect.~\ref{sec:projection}, with $V_{ik}$ obtained from Eq. \ref{eq:VrhoMaxrhoMinRmaxRmin} with the two cylindrical radii correspond to the index $k$, and the two spherical radii to the index $i$.
Given that the radial profiles are given as a function of $r/r_{200}$, the volume in Eq. \ref{eq:Pdiscrete} must be multiplied by the value of $r_{200}^3$ in cm$^3$ to obtain the luminosity in cgs units.

\subsection{The PyAtomDB code}

This project makes use of the \texttt{AtomDB} data and models of X-ray and extreme ultraviolet emitting astrophysical spectra for hot and optically thin plasma. These data were originally developed by \cite{smith2001} as the Astrophysical Plasma Emission Database (\texttt{APED}) and the Astrophysical Plasma Emission Code (\texttt{APEC}). These codes result
from a development started with  \cite{raymond1977} and was followed by the codes of \cite{mewe1985,mewe1986} and \cite{kaastra1992}. Over the years, as new and improved calculations in atomic data became available, \texttt{AtomDB} was improved by \cite{foster2012} and eventually led to the  \texttt{PyAtomDB} project that is easily accessible through \texttt{python} \citep{foster2020}. The latter is the version that we use in this project.

The \texttt{AtomDB} code provides the emissivity of a plasma in collisional ionization equilibrium (CIE, e.g., \citealt{mazzotta1998}), using the \cite{anders1989} Solar abundances that can be re--scaled, either individually or as a whole, by an overall
metal abundance $A$ as a fraction of Solar. Moreover, the emissivity can be conveniently decomposed into
the contribution from each element; and, for each element, it is possible to calculate the contribution from each
transition, e.g., it is possible to calculate the power of the \ovii\ resonance line at 21.602~\AA, or that
of any other X--ray line of interest. 

\subsection{Mean atomic weights of the plasma}

Given that the emissivity calculated via \texttt{atomDB} is rescaled by the product $(n_e \cdot n_H)$ of the plasma, which is a function of
the metal abundances, it is also necessary to specify the mean weights of the various plasma species.
In general, the number density of species $j$ is related to the total mass density of the medium $\rho$ via
\begin{equation}
  n_j =\dfrac{\rho\, X_j}{m_j}
\end{equation}
where $X_j$ is the \textit{fraction by mass} of the $j$--th species (say, H, He, electrons, heavy ions, etc.),
\[
  X_j = \dfrac{\rho_j}{\rho},
\]
and $m_j$ is the atomic mass of the species ($\sim m_p$ for hydrogen, $\sim 4 m_p$ for helium, etc.).
The mass density $\rho$ (in a given shell) includes all species that are present in the plasma,
\begin{equation}
\begin{aligned}
  \rho =&\rho_H + \rho_{He} + \rho_{z} \simeq \\
  &  n_H m_p + n_{He}\, (4 m_p) + \left(\sum_{z} n_z \cdot m_z \right) 
  .
  \end{aligned}
\end{equation}
where $z$ labels all metals, defined as elements with atomic number $Z>2$.
It is customary to define
the hydrogen mass fraction as $X$, that of helium as $Y$, and that of all other metals combined
as $Z$ (not to be confused with the same letter that is also used for the overall metal abundance).  It is then true that the sum of the mass fractions is
\begin{equation}
  X+Y+Z=1.
  \label{eq:XYZ}
\end{equation}
According to typical Solar abundance ratios, the number density of He is
approximately 10\% of that of H. Specifically,  $n_{He} \simeq 0.1 \cdot n_{H}$ according to \cite{anders1989}, 
while \cite{lodders2003} finds a number ratio of $n_{He}=0.084 \cdot n_H$.
Assuming that $Z=0$ (negligible contribution from metals to mass), and an 8.4\% ratio by number of helium-to-hydrogen,
the mass fractions are approximately $X=0.749$ and $Y=0.251$.

It is now convenient to define the \textit{mean molecular weights} of a given species via
\[
  \begin{cases}
    \mu_e = \dfrac{\rho}{m_p n_e} \;\; \\[10pt]
    \mu_H = \dfrac{\rho}{m_p n_H} \;\;\\[10pt]
    \mu_{He} = \dfrac{\rho}{m_p n_{He}} \\[10pt]
    \mu = \dfrac{\rho}{n \; m_p}\;\;
  \end{cases} 
\]
In particular, the mean molecular or atomic weight $\mu$ is the average weight (in units of the proton mass $m_p$)
of a given particle using the density of all particles, $n=n_e+n_H+n_{He}+n_{x}$. It therefore
follows that
\begin{equation}
  \dfrac{1}{\mu} = \dfrac{1}{\mu_e} + \dfrac{1}{\mu_H}+  \dfrac{1}{\mu_{He}} \left(+ \dots + \dfrac{1}{\mu_z} + \dots\right)
  \label{eq:mus}
\end{equation}
where the other terms can be ignored inasmuch as there is a small fraction of metals.
Clearly \eqref{eq:XYZ} and \eqref{eq:mus} are equivalent.

Characteristic values for the mean atomic weights in the ICM, assuming fully ionized plasma with primordial
abundances (metal mass fraction $Z=0$), are $\mu_H = 1.335$, $\mu_e=1.143$, $\mu_{He} = 15.94$ and $\mu = 0.593$.
%
This means that 1 electron traces an equivalent mass of 1.143 protons, and 1 hydrogen atom traces approximately 
the mass of 1.335 protons.  When metals are taking into account, their values are slightly larger,
while the ratio  $\mu_H/\mu_e =n_e/n_H \simeq 1.168$ remains approximately the same. From the \tng\ simulations
we measure a ratio  of approximately $n_e/n_H =1.16$ for all species, at all
radii, which is consistent with these calculations.

Collisional ionization equilibrium (CIE) is believed to apply in the hot ICM, and with good approximation
also to the lower density warm--hot phase at $\log T(\mathrm{K}) =5-7$. Utilizing the CIE approximation, it is 
therefore also possible to predict column densities of any ion of interest along a sightline 
at any projected radius from the center of the cluster, and for any phase. This gives us the ability to
study the  detectability of absorption lines from a give atomic species. 
In this initial study, we focus on four ions that are
prominent in CIE at the temperatures of the warm--hot phase, namely \ovii, \oviii, \neix, and \nex \citep[e.g.][]{mazzotta1998}, and probe both their soft X--ray emission and their column densities. A more detail analysis on their detectability (through current and upcoming X--ray instruments) will be presented in a following paper. Table~\ref{tab:lines} provides the atomic data for the lines of interest.

\begin{table*}[]
    \centering
    \begin{tabular}{l|llllll}
    \hline
    \hline
    Ion & Transition & Line Name & Wavelength (\AA) & Osc. Strength $f$ & Upper level & Lower level \\[5pt]
    \hline
             \ovii & $1s^2-1s2p$ & He$\alpha$ & 21.602 & 0.696  &  7 & 1\\
         \oviii & $1s-2p$ & Ly$\alpha$& 18.969 (18.973,18.967) & 0.416 & 3, 4 & 1\\
         \neix & $1s^2-1s2p$ &  He$\alpha$ & 13.447 & 0.724 & 7 & 1\\
         \nex & $1s-2p$ & Ly$\alpha$ & 12.134 (12.138,12.132) & 0.416  & 3, 4& 1\\
         \hline
         \hline
    \end{tabular}
    \caption{Atomic data for selected X--ray lines of interest, for ions that are expected to be abundant in the warm--hot phase. Upper and lower levels are used for line identification in the \texttt{APEC} database.}
    \label{tab:lines}
\end{table*}

\subsection{The soft X--ray emission near clusters : Initial results}

\label{sec:emission}
\begin{figure}
    \centering
\includegraphics[width=3.5in]{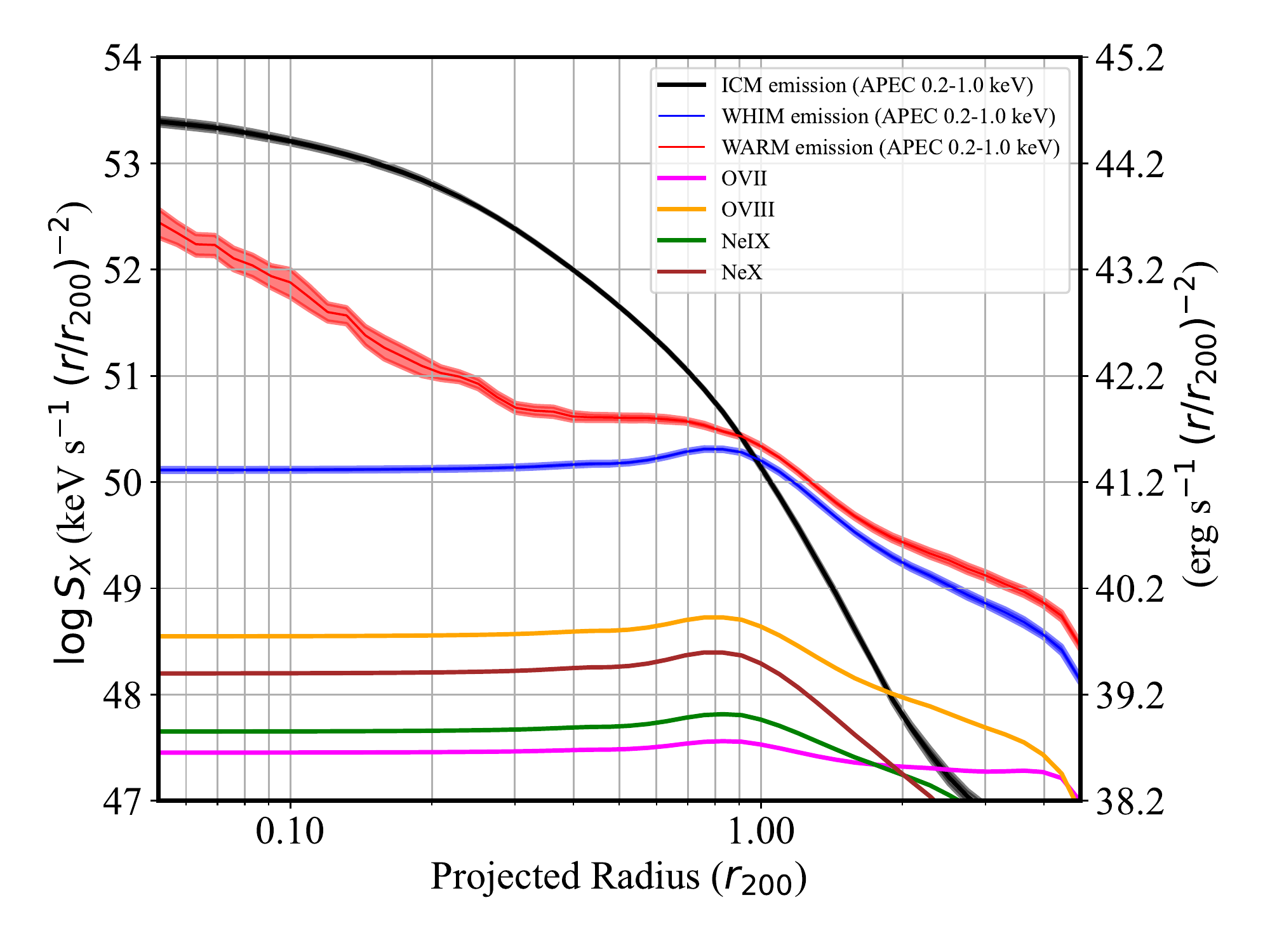}
    \caption{The X-ray Surface Brightness $S_X$ (Luminosity per unit area) in a soft X--ray band of choice (0.2-1~keV) for the various phases of the gas, for all clusters. }
    \label{fig:Lx}
\end{figure}

\begin{figure}
    \centering
    \includegraphics[width=3.5in]{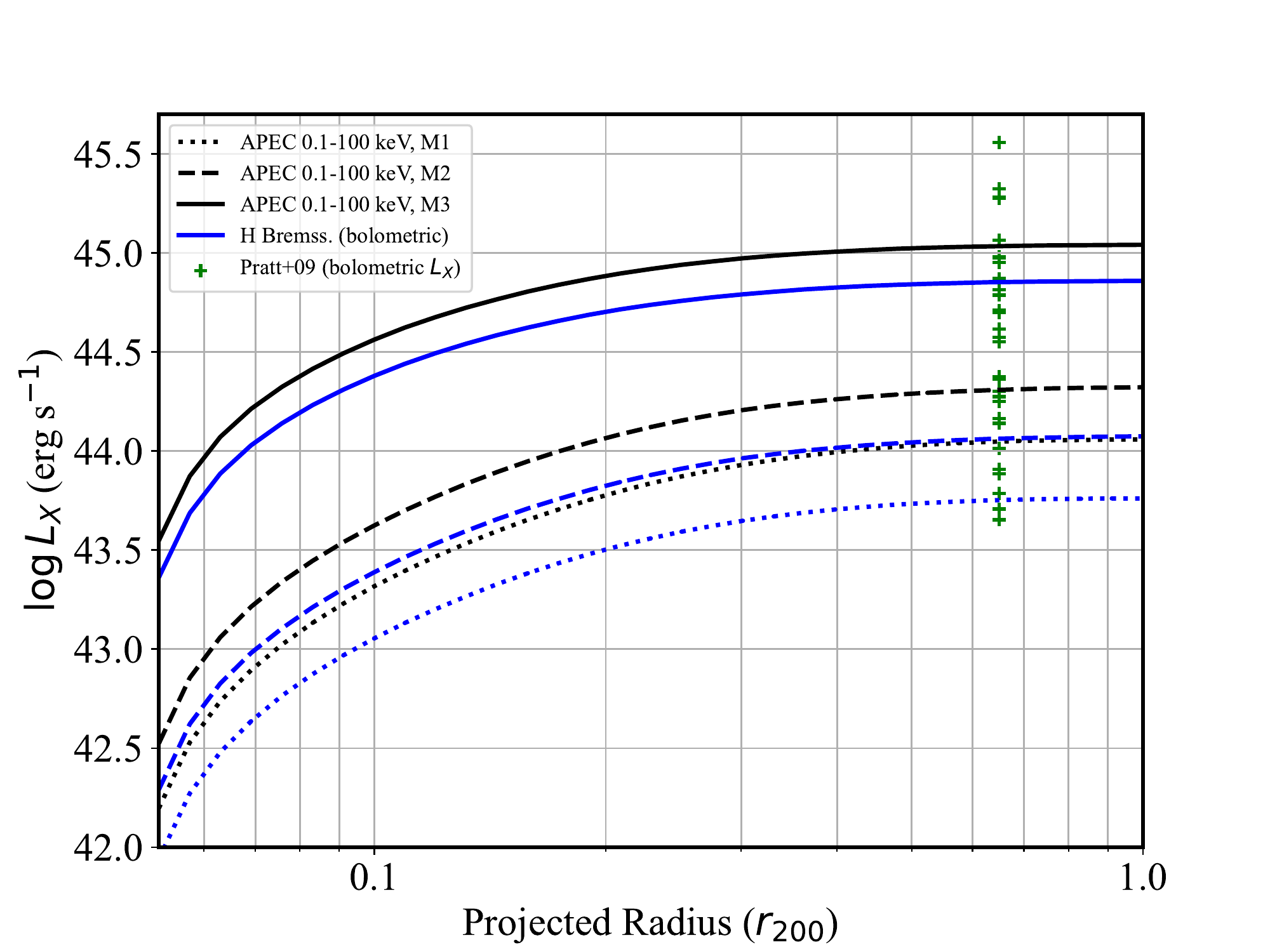}
    \caption{The X-ray Luminosity in a broad X--ray band (0.1-100~keV) for the ICM phase, for the three cluster mass bins: $M_1$ (dotted lines), $M_2$ (dashed lines) and $M_3$ (solid lines).
    This broad--band luminosity is compared with bolometric luminosities of massive clusters from \cite{pratt2009}}
    \label{fig:LxBol}
\end{figure}

We use Eq.~\ref{eq:Pdiscrete} to measure the average luminosity (proportional to the count rate) of the three phases as a function of  radial distance from the center of a cluster.
Given that the radial profiles are normalized to the $r_{200}$ of each cluster in the \tng\ simulations, the volumes
$V_{i,k}$ obtained from Eq.~\ref{eq:VrhoMaxrhoMinRmaxRmin} are multiplied by $r_{200}^3$ in units of cm$^{3}$.

Figure~\ref{fig:Lx} features the luminosity per unit area, or surface brightness
according to Eq.~\ref{eq:Sx} in a soft X--ray band of choice, as a function of the projected radius,
for all the clusters.
As illustrated in Figure~\ref{fig:Lx}, it is clear that just outside of the virial radius, the luminosity of the warm gas becomes dominant, relative to that of the hot ICM. Inside clusters ($<1 \ R_{200}$), the ICM gas dominates the soft X--ray surface brightness.
Moreover, the WHIM gas has a very low luminosity inside clusters, 
while the entire WARM phase has a $\geq 1$~\% surface brightness
compared to the hot ICM. As discussed previously in section~\ref{sec:TNG}, we expect that the soft X--luminosity of warm gas inside clusters is mostly the result of WCGM gas clumps, whereas outside clusters it is mostly induced by WHIM gas. Indeed, beyond $\ 1 \ R_{200}$, the soft X--ray luminosity of WHIM gas becomes higher than ICM luminosity. 

In Fig.~\ref{fig:Lx}, we also highlight the contribution of the total emission by selected ions (\ovii, \oviii, \neix\ and \nex), inclusive of their continuum and line emission. 
Focusing on the metal-line emission, we found that surface brightness of metals is in general decreasing beyond the virial radius, with \oviii\ the brightest line followed by \nex, in agreement with previous predictions with OWLS simulations \citep[see Fig. 1 of][]{VanDeVoort2013}. Moreover, our emission of \ovii\ is also consistent with \cite{Tuominen2023} who found that \ovii\ drops dramatically beyond the virial radii of halos in the \texttt{EAGLE} simulations. 
In detail, beyond the virial radius of massive clusters, the 
distribution and oxygen  ions (primarily \ovi\ and \ovii) are expected to follow cosmic filaments \citep[as described in][]{Artale2021,Tuominen2023}, and thus, oxygen ions might be good tracers of WHIM gas in filaments connected to clusters \citep{gouin2022}. In fact, \cite{nelson2018} have used \tng\ simulation to show that \ovi, \ovii, and \oviii\ are mostly induced by the WHIM gas phase. Further details on ions abundances and their emissions as a function of their location (e.g., inside or outside filaments) and gas phases will be discussed in detail in another publication. 

In order to address the agreement with measurements of the X--ray luminosity from the hot ICM, Fig.~\ref{fig:LxBol} shows a bolometric luminosity in a broad X--ray band (0.1-100 keV; black lines) and the bolometric contribution from free--free bremsstrahlung emission from hydrogen alone, for the three cluster mass bins, according to
\begin{equation}
    \overline{\epsilon}_{\text{ff}} = 1.4 \times 10^{-27} n_e n_H T^{1/2} \overline{g}_B \; (\text{erg s}^{-1}\text{cm}^{-3})
\end{equation}
where $\overline{g}_B \simeq 1.2$ is the velocity and frequency--averaged Gaunt factor \citep[e.g.,][]{rybicki1979}. 
In Fig.~\ref{fig:LxBol}, the bolometric X--ray luminosities from the \texttt{REXCESS} sample of massive clusters out to $r_{500} \simeq 0.65 \, r_{200}$ are shown as green crosses \citep{pratt2009}. 
This figure shows that their measurements are consistent with the \tng\ simulated clusters, which spread over a range of cluster masses from $10^{14}$ to $10^{15} M_{\odot/h}$.  

\subsection{The soft X--ray emission near clusters due to WARM gas}

\begin{figure*}[!h]
    \centering
\includegraphics[width=3.5in]{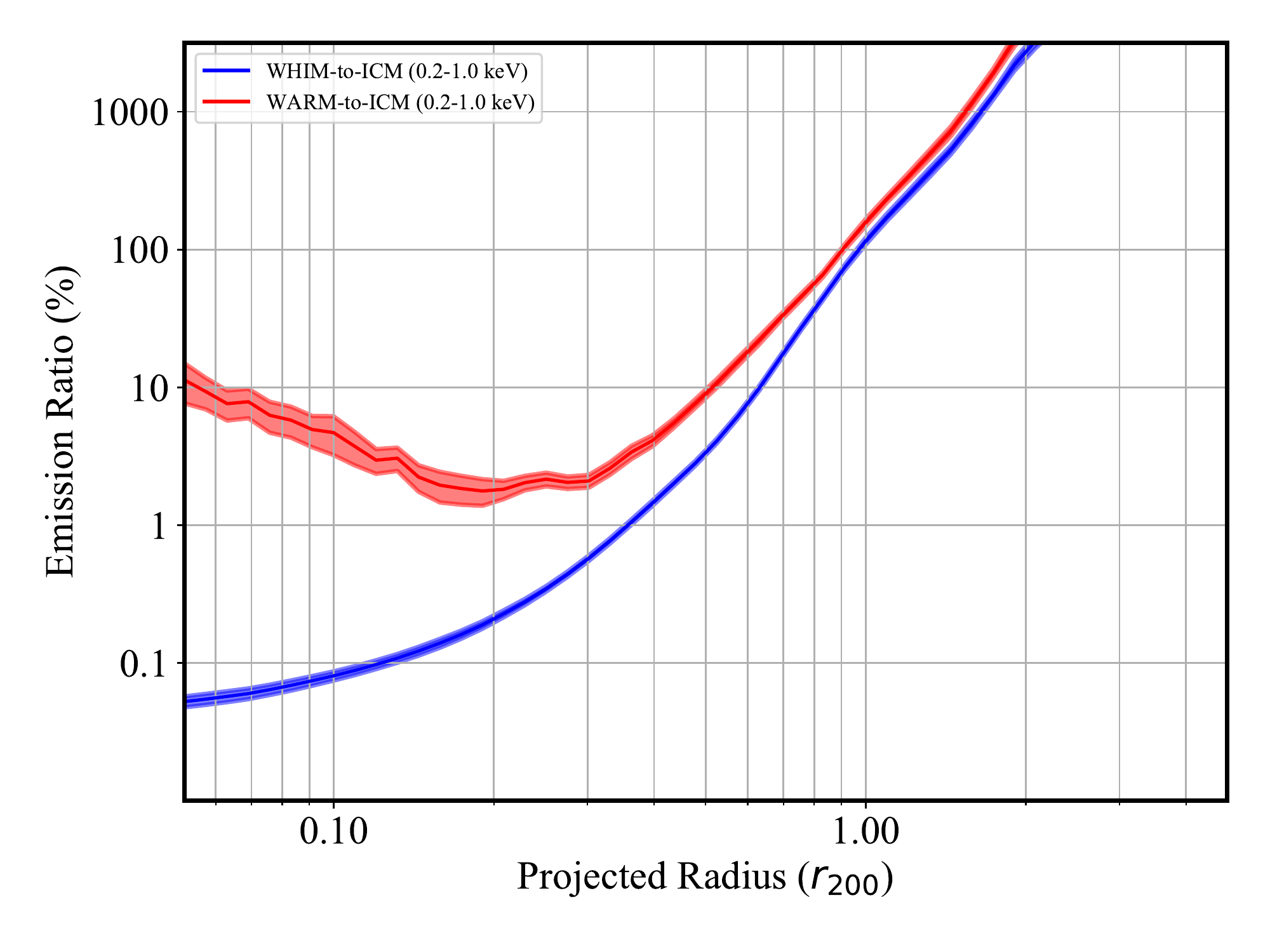}    
\includegraphics[width=3.5in]{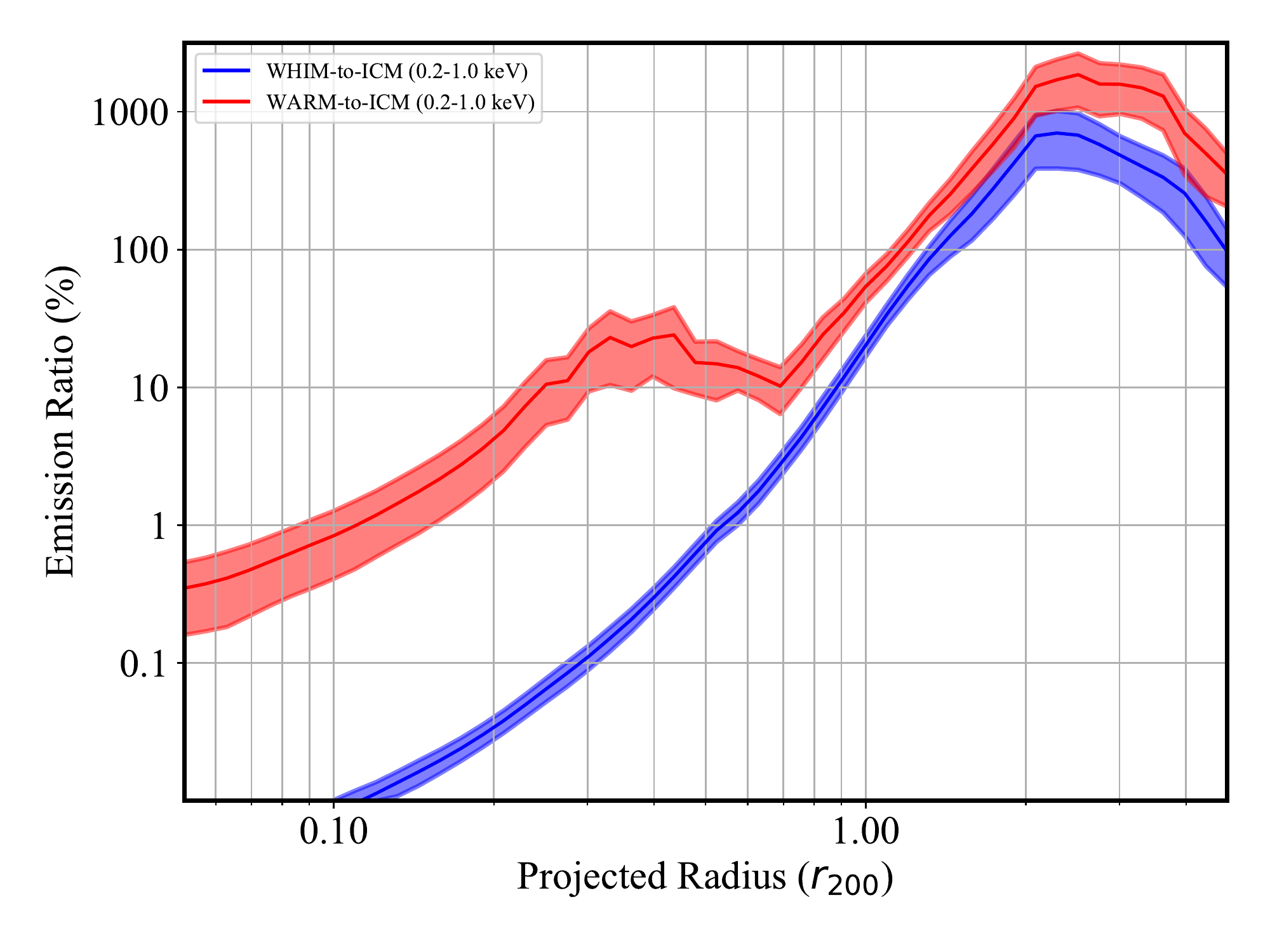}
    \caption{Ratio of luminosities or count rates of warm phases, relative to the hot ICM, left for all clusters, and
    right for mass sample $M_3$. The figure also shows the contribution of individual ions, primarily through their line emission, to the total emission from warm subvirial phases. 
    \label{fig:LxRatio}}
\end{figure*}

Figure~\ref{fig:LxRatio} shows the ratio of WHIM--to--ICM and WARM--to--ICM phases, for the same soft X--ray band as in Fig.~\ref{fig:Lx}. This result allows us to highlight the relative importance of the WARM and WHIM gas in the X-ray emission of circum–cluster gas, relative to the usual ICM gas.
These ratios are consistent with an increasing radial trend in the relative count rate from warm sub-virial gas, relative to the virialised ICM, which reaches a ratio of 100\% just outside the virial radius. 
Moreover, the high--density WARM gas is also capable of producing around $10$\% of soft X--ray excess emission, compared with that from the hot ICM, towards the interior of the cluster. 
For most massive clusters ($M_3$ mass bin), in the left panel Fig.~\ref{fig:LxRatio}, the ratio WARM--to--ICM is constant at a value of 10\% from radius $0.2 r_{200}$ up to $0.7\times r_{200}$.
These results are quantitatively consistent with the presence of soft X--ray excess emission that was detected by ROSAT and other instrument towards Coma out to the virial radius\citep[e.g.][]{bonamente2003,bonamente2023}, and in the inner regions of several other clusters \citep[e.g.][]{bonamente2002}.
An in--depth analysis of the soft excess emission from clusters, which was originally discovered in extreme--ultraviolet observations of Coma and the Virgo clusters \citep{lieu1996a,lieu1996b} will be presented in a follow--up paper. 

\begin{figure}[!h]
    \centering
    \includegraphics[width=3.5in]{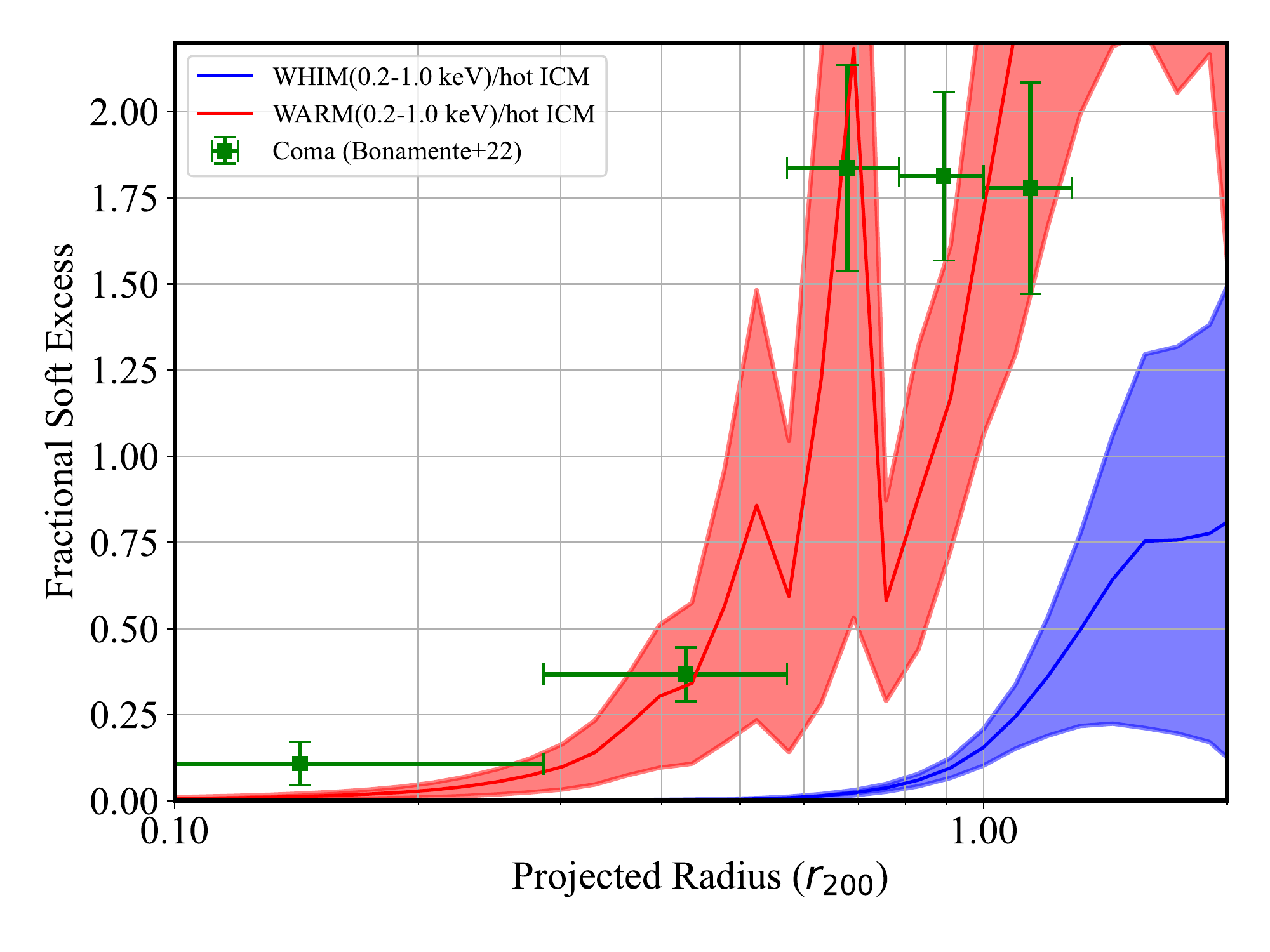}
    \caption{Ratio of count rates from the warm-hot phases, relative to the hot ICM and same as in Fig.~\ref{fig:LxRatio}, but for the most massive clusters at $M_{200} >10^{14.75}$ M$_{\odot}/h$) from sample M4,
    to match the mass of the Coma cluster. The data points are the soft excess emission
    measured by \cite{bonamente2022c} in the 0.2-1.0 keV band, based on the ROSAT analysis
    of \cite{bonamente2003}.
    }
    \label{fig:comaSoftExcess}
\end{figure}

A comparison of the \cite{bonamente2022c} soft excess emission in Coma with the \tng\ prediction is illustrated in Fig.~\ref{fig:comaSoftExcess}, where only the most massive clusters ($M_{200} > 10^{14.75} M_{\odot/h}$) are used for the prediction, to match Coma's virial mass as measured by \cite{mirakhor2020} as $M(r_{200})=(8.50\pm0.55) \times 10^{14}$ M$_{\odot}$. One can see that our numerical prediction of soft X--ray excess via WARM gas emission (red line), is in very good agreement with Coma soft excess (green data points). This emission from warm--hot gas near the virial radius matches the level of soft excess emission detected with ROSAT, indicating a thermal origin for the emission, above the contribution from the hot ICM.
Notice that the large scatter of ratio profile, especially from $0.5$ to $0.9 \ r_{200}$, is due to the presence or absence of WARM gas clumps (from one cluster to another), which drastically increase the soft excess, and thus, induce this large scatter traced by the errorbars.

Similar studies were recently performed with the \texttt{Magneticum} simulation \citep{Churazov2023}, finding that the emission from the diffuse WHIM phase alone was not sufficient to explain the observational result of \cite{bonamente2022c}. This is also clearly shown by our results when only the WHIM gas is considered (blue curve in Fig.~\ref{fig:comaSoftExcess}). A qualitatively similar result was also obtained by the \cite{cheng2005} simulations, which identified
low--entropy and high--density warm gas as the likely phase responsible for the soft excess within the virial radius.


\subsection{Soft X--ray absorption lines : Column density of ions}

\begin{figure}
    \centering
\includegraphics[width=3.5in]{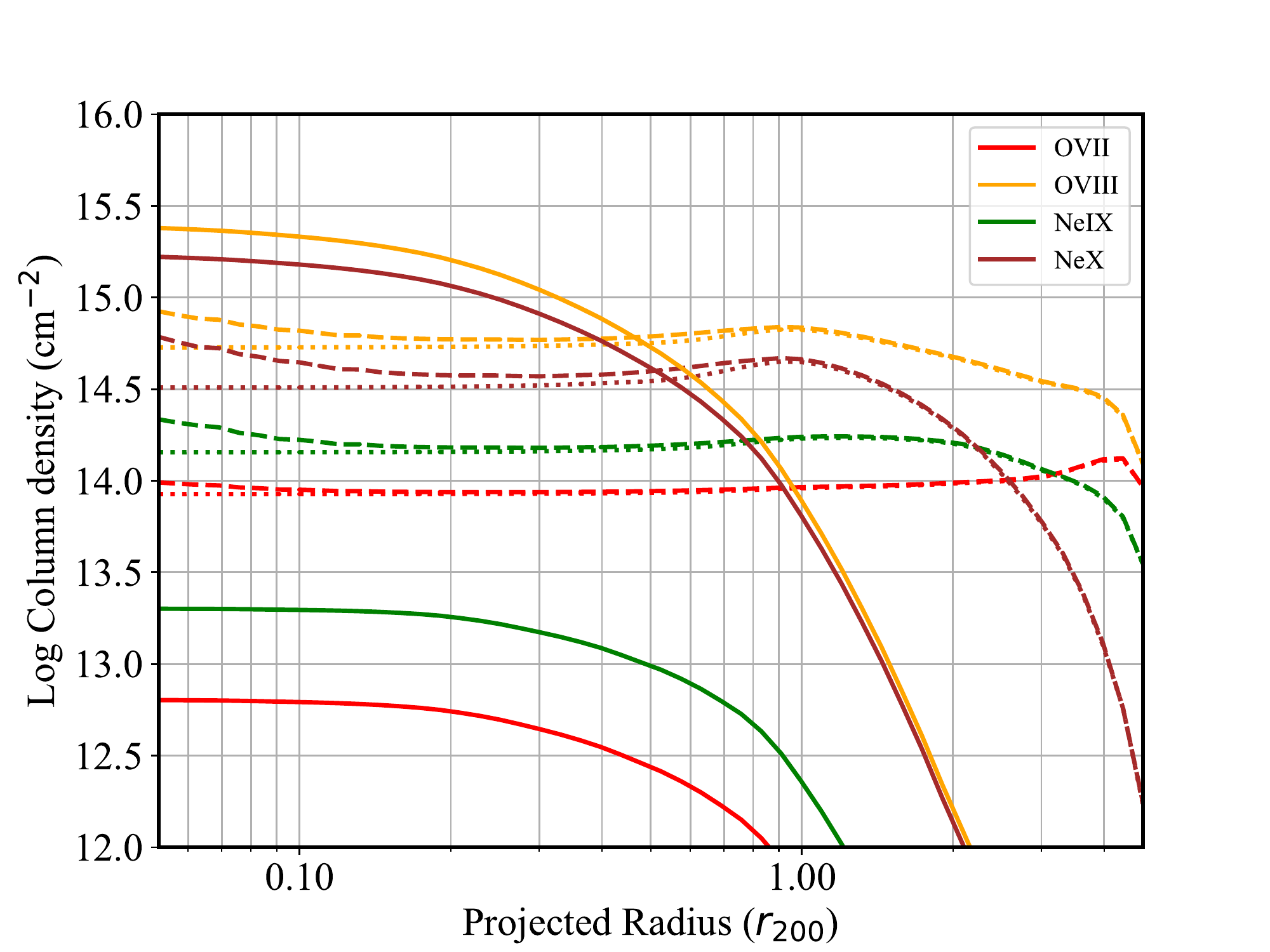}
    \caption{Projected column densities of selected X--ray ions for all the clusters. Solid curves are for the HOT phase, dashed curves for the WARM and dotted curves for the WHIM phases.}
    \label{fig:ColDen}
\end{figure}

Our analysis of the \tng\ simulations also measures the column density of any ion of interest, which can be used to address the feasibility of
detecting the warm sub--virial gas in absorption. 
For example, we illustrate in Fig.~\ref{fig:ColDen} the average projected column density of four ions of interest for X--ray absorption studies, viz. \ovii, \oviii, \neix\ and \nex.
There have been several reports of possible absorption lines associated from intervening warm--hot gas \citep[recently, e.g., ][]{bonamente2016,kovacs2019,ahoranta2020,ahoranta2021,spence2023}, although usually with limited significance.
In particular, given the lower chance alignment of a bright background extragalactic source (e.g., a quasar) with
a galaxy cluster along our sightline, there have been to date no attempts to detect the warm--hot gas in absorption near clusters. 
Our preliminary results of Fig.~\ref{fig:ColDen} indicate that there is especially a substantial amount of \oviii\ that is in fact within the reach of upcoming X--ray missions such as the proposed \textit{Light Element Mapper (LEM)} mission \citep{LEM2022}. 

These predictions of X-ray absorption lines in circum–cluster environments are done with \tng\ simulation, which already show a good agreement with observations of \ovi\ column density at low redshift \citep{nelson2018}. 
Moreover, as recently highlighted by \cite{butler2023} using \texttt{CAMELS} simulations, \ovii\ absorption predictions are quite sensitive to change with SN feedback models. Qualitatively, our prediction of \ovii\ column density is in agreement with \cite{butler2023} who found that the OVII absorbers with column densities $\log N_{\mathrm{OVII}} <14 \ \mathrm{cm}^{-2}$ fall within $1$ Mpc from the center of halos. They thus predicted that the simulated \ovii\ column density are higher in halos and their outskirts than in the large-scale cosmic web.
Along this line, the \texttt{EAGLE} simulations analyzed by \cite{wijers2022}
estimated the detectability of X-ray metal-line emission from intracluster gas, and found that \ovii\ and \ovii\ emission lines will be detectable out to the virial radius by the proposed \texttt{Athena} X-IFU instrument.
These initial results of X--ray absorption lines from the different gas phase in circum–cluster environments illustrate an example of numerical predictions done through our methodology (Sect.~\ref{sec:SX}), and will be
 explored in more detail a subsequent publication, including the effects of gas clumping and the volume covering fraction.

\section{Discussion and conclusions}
\label{sec:conclusions}

This paper has presented the methods of analysis for the thermodynamic quantities of simulated galaxy clusters from  \tng\ simulation (with  $M_{200}>10^{14} M_{\odot}/h$ at $z=0$), for the purpose of studying the X--ray emission in the circum--cluster environment. Given the wealth of information provided 
by these simulations, 
this paper is intended to be the first in a series of publications where we will analyze the properties of sub--virial gas in and around galaxy clusters, with emphasis on its soft X--ray emission and the effect of the emission on related cluster observables.

Starting from this simulated cluster sample, we extracted azimuthally--averaged radial profiles of the density, temperature and metallicity for different gas phases \citep[e.g.][]{martizzi2019,Galarraga21}: the hot ICM phase at $\log T(K)>7$ and the WARM phase at $5 \leq \log T(K) \leq 7$. The latter was further divided into WHIM ($\log n_H(\text{cm}^{-3}) \leq -4$) and WCGM gas, having higher densities.
The goal was to provide a simple yet accurate set of averaged quantities that are representative of the hot virialized gas and of the cooler sub--virial gas that is preferentially found in either filamentary structures (WHIM), or in higher--density regions such as the surroundings of haloes (WCGM). 
Indeed, based on this same simulated cluster sample, \cite{gouin2022} have found that gas in the circum--cluster environment is at the interface between the warm gas infalling in filaments connected to clusters, and the hot virialized gas inside clusters \citep[see also][]{Vurm2023}.

A first step in our analysis was to validate the \tng\  data with results from observations of galaxy clusters. Temperature profiles of the hot ICM are well studied in X--rays, and the \tng\ profiles (see Sect.~\ref{sec:profiles} and Fig.~\ref{fig:T}) are in excellent agreement with the X--ray measurements of a large sample of X--ray clusters of \cite{ghirardini2019}, and with the \cite{mirakhor2020} radial profile of the Coma cluster, one of the nearest and most massive clusters.
A similar agreement applies to the measurement of the chemical abundance of the ICM (see Fig.~\ref{fig:A}), especially towards the virial radius where the abundance of the gas reaches approximately a value of 20\% Solar. 
Another key factor for the prediction of X--ray emission is the amount of inhomogenieties in the plasma, since the emissivity is proportional to the square of the plasma density. 
Accordingly, we have measured the clumpiness parameter (see Sec.~\ref{sec:clumpiness}), and, as expected, we found that the inhomogenieties is increasing with the radial distance from clusters \citep{nagai2011,Roncarelli2013,angelinelli2021}. Our clumpiness radial profile is moreover in good agreement with X-COP clusters \citep{eckert2015}, suggesting a good modeling of AGN feedback, a key ingredient for the study of gas inhomogeneities \citep{Planelles2017}. Interestingly, separating gas in various phases, we show that both the ICM, WARM, and WHIM gas phases have significant different level of inhomogenieties, within and beyond the virial radius (see Fig.~\ref{fig:clumping}).

To study the X--ray radiation from the various plasma phases, we devised a simple analytical method to project the radial profiles of the thermodynamical quantities of interest. 
This projection, in conjunction with the X-ray emissivity codes and databases provided by the \texttt{pyatomdb} project \citep[e.g.][]{foster2020}, makes it possible to predict the X--ray luminosity and surface brightness profiles projected on a detector plane. 
A key test for the reliability of our \tng\ data is the ability to reproduce the X--ray luminosity of the ICM of observed clusters (see Fig.~\ref{fig:LxBol}), with bolometric luminosities in the range
$\log L_X(\text{erg/s)}) = 44-45$, similar to that of observed clusters \citep[e.g.][]{pratt2009}. 

This paper has also presented initial results for the analysis of the X--ray emission from the warm sub--virial gas. To date, there are no significant observational constraints on soft X--ray emission from circum--cluster sub--virial plasma, due to the challenges associated with the detection of X-rays in emission at those temperatures. 
The preliminary analysis of Sect.~\ref{sec:SX} shows a fundamental new result that is the key take--away from this initial paper: the warm subvirial gas, either in the form of low--density WHIM gas or as higher--density WCGM,  features a surface brightness that is at the same level as that from the ICM near the virial radius, and it becomes the dominant source of soft X--ray radiation at larger distances from the center. This is a key new result that had not been previously appreciated, and it will be the basis for a more detailed investigation in an separate publication.

For example, Fig.~\ref{fig:LxRatio} shows that the average soft X--ray emission from the WARM gas in entire cluster sample reaches 100\% of the ICM emission at $r/r_{200} \simeq 0.8-1$ (left panel), with  a similar result for the most massive clusters alone (right panel). It is therefore clear that ignoring this emission component would lead to biases in the inference of such quantities as cluster masses measured from X--rays.
In the inner cluster regions, the relative importance of the soft X--ray emission from  warm gas is usually at a lower level of $\sim$1-10\%.
In particular, focusing on the Coma cluster, and its soft X--ray emission \citep{bonamente2023}, we found that the most massive \tng\ clusters reproduce well Coma's soft X--ray excess (see Fig.~\ref{fig:comaSoftExcess}), supporting our finding that WARM gas is a crucial ingredient to understand soft X--ray emission in circum–cluster environments.

In summary, the preliminary results presented in this paper
are qualitatively consistent with a thermal origin for the
cluster soft excess emission detected in a number of clusters, such as in Coma \citep{bonamente2003,bonamente2022c} and in other clusters \citep{bonamente2002}. 
The emission, which is at the $\sim 10$\% level in the inner regions
and rising in importance towards the virial radius, is naturally explained by the warm--hot subvirial phases seen in these \tng\ simulations.
A more detailed analysis of the soft X--ray emission is deferred to 
a follow--up publication, where we also plan to present results 
as a function of cluster masses and other structural properties, such as the degree of connectivity that is used as a proxy for the possible presence of WHIM filaments \citep{gouin2022}, whether such emission is due to the denser WCGM or to lower--density WHIM filaments, and its dependence on cluster properties.

\begin{acknowledgements}
This research has been supported by the funding for the ByoPiC project from the European Research Council (ERC) under the European Union’s Horizon 2020 research and innovation program grant agreement ERC-2015-AdG 695561 (ByoPiC, https://byopic.eu). 
CG thanks the very useful comments and discussions with all the members of the ByoPiC team, and in particular N. Aghanim.
We thank the IllustrisTNG collaboration for providing free access to the data used in this work. 
CG is supported by a KIAS Individual Grant (PG085001) at Korea Institute for Advanced Study.
\end{acknowledgements}

\bibliography{max}
\bibliographystyle{aa}

\appendix

\section{Statistical analysis of the radial profiles}
\label{Appendix:Stat}

The radial profiles of the thermodynamic quantities (Figures~\ref{fig:ElDens} through \ref{fig:clumping})
are obtained by averaging the quantity of interest, for each of the \nclu\ clusters at each  of the \nrad\ radii.
We then obtain a distribution of the quantity, for all clusters that have data at that radius. 
At a given radius, not all of the clusters have data, and therefore certain data points --- such as 
those at the lowest radii for the electron density of the WARM phase in Fig.~\ref{fig:ElDens}, bottom--right panel --- result from the use of only a small fraction of all the clusters in the sample. When fewer than 3 clusters have data, we chose not to report that quantity.

\begin{figure}[!h]
\centering
    \includegraphics[width=2.8in]{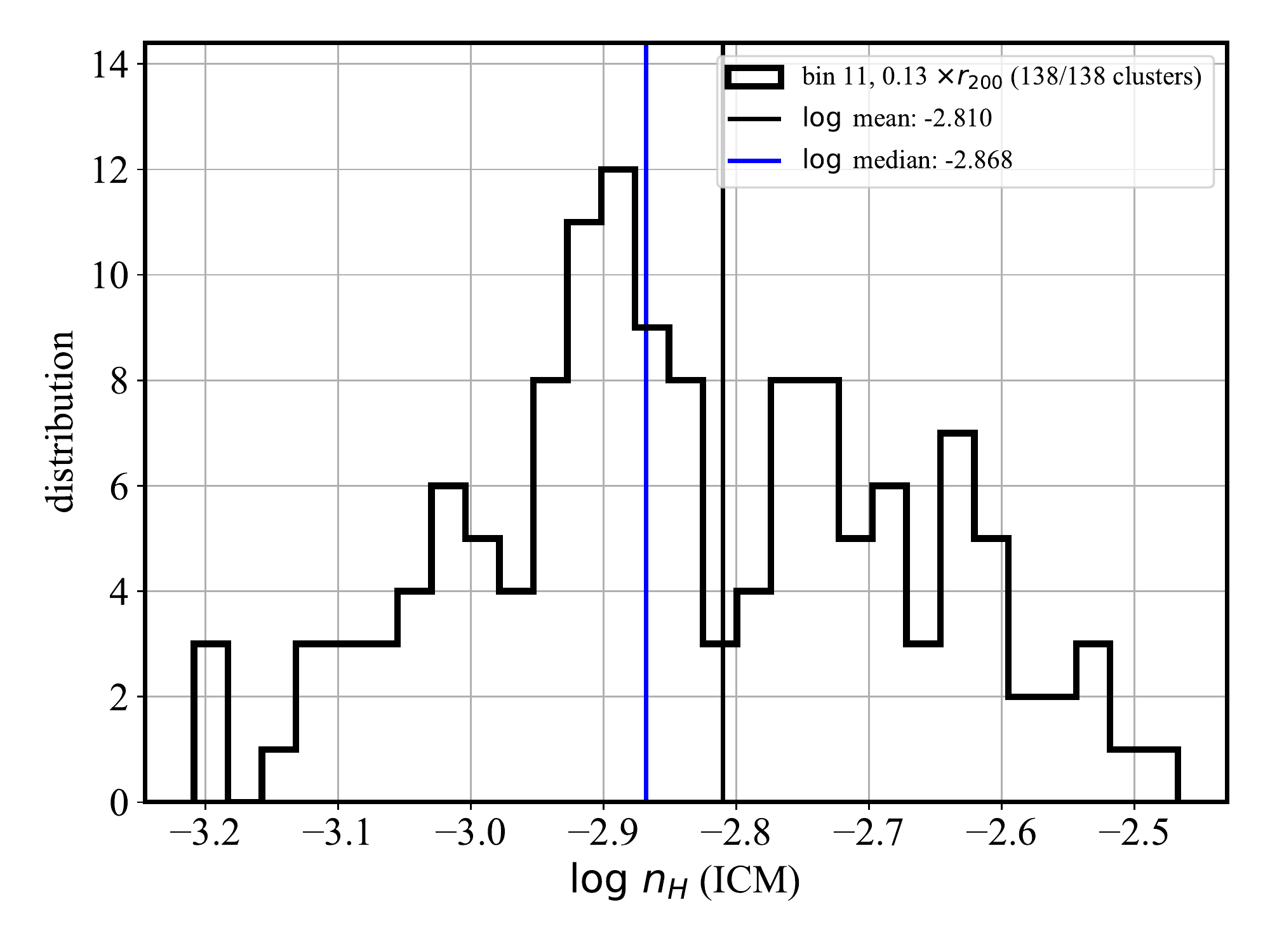}
    \includegraphics[width=2.8in]{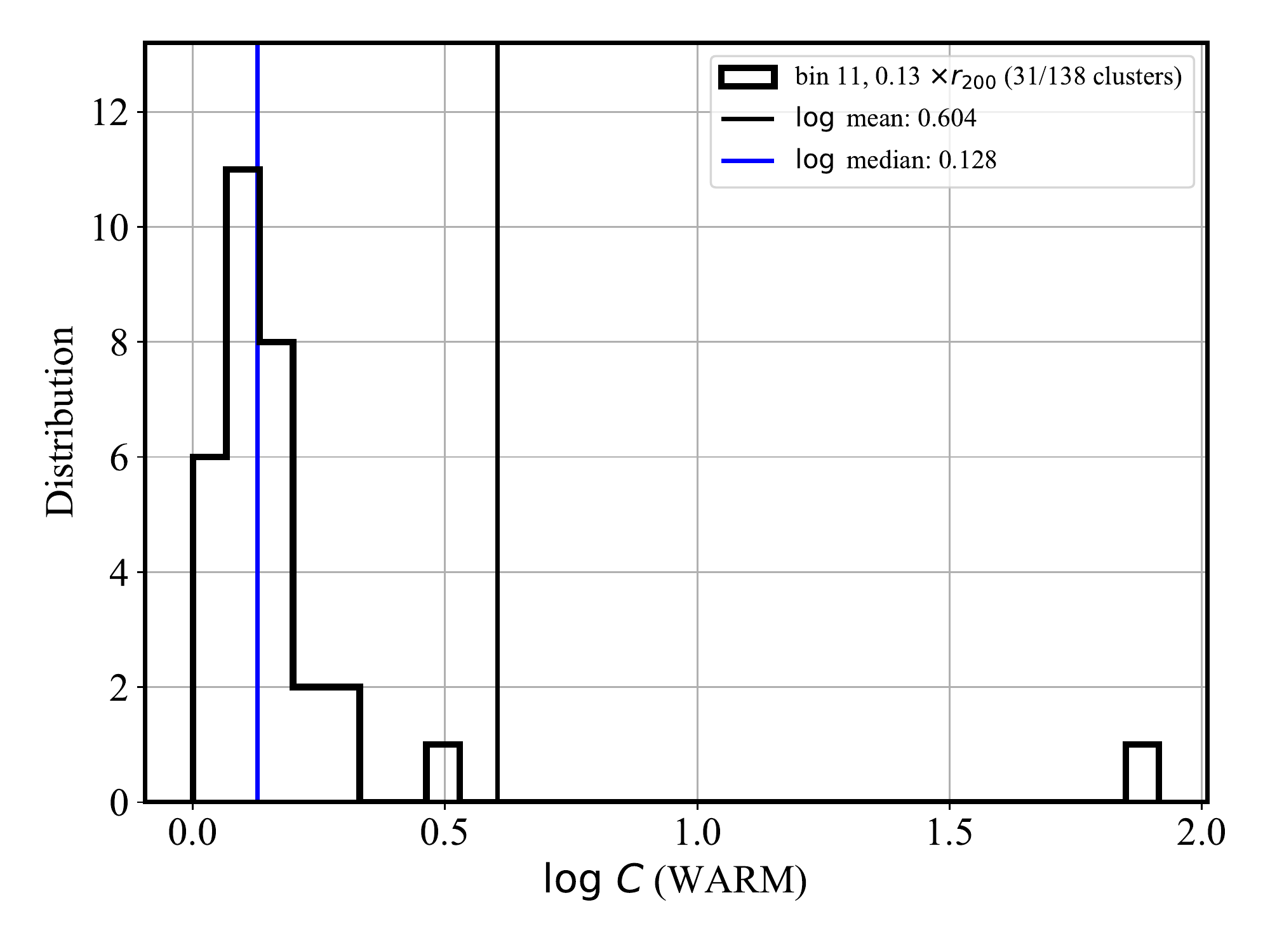}
    \caption{Distribution of values of selected quantities from the analysis of the \nclu\ \tng\ clusters, for the radial bin 11. Top:
    ICM density; Bottom: clumpiness factor for WARM phase.}
    \label{fig:distr}
\end{figure}

\begin{figure}[!h]
\centering
    \includegraphics[width=2.8in]{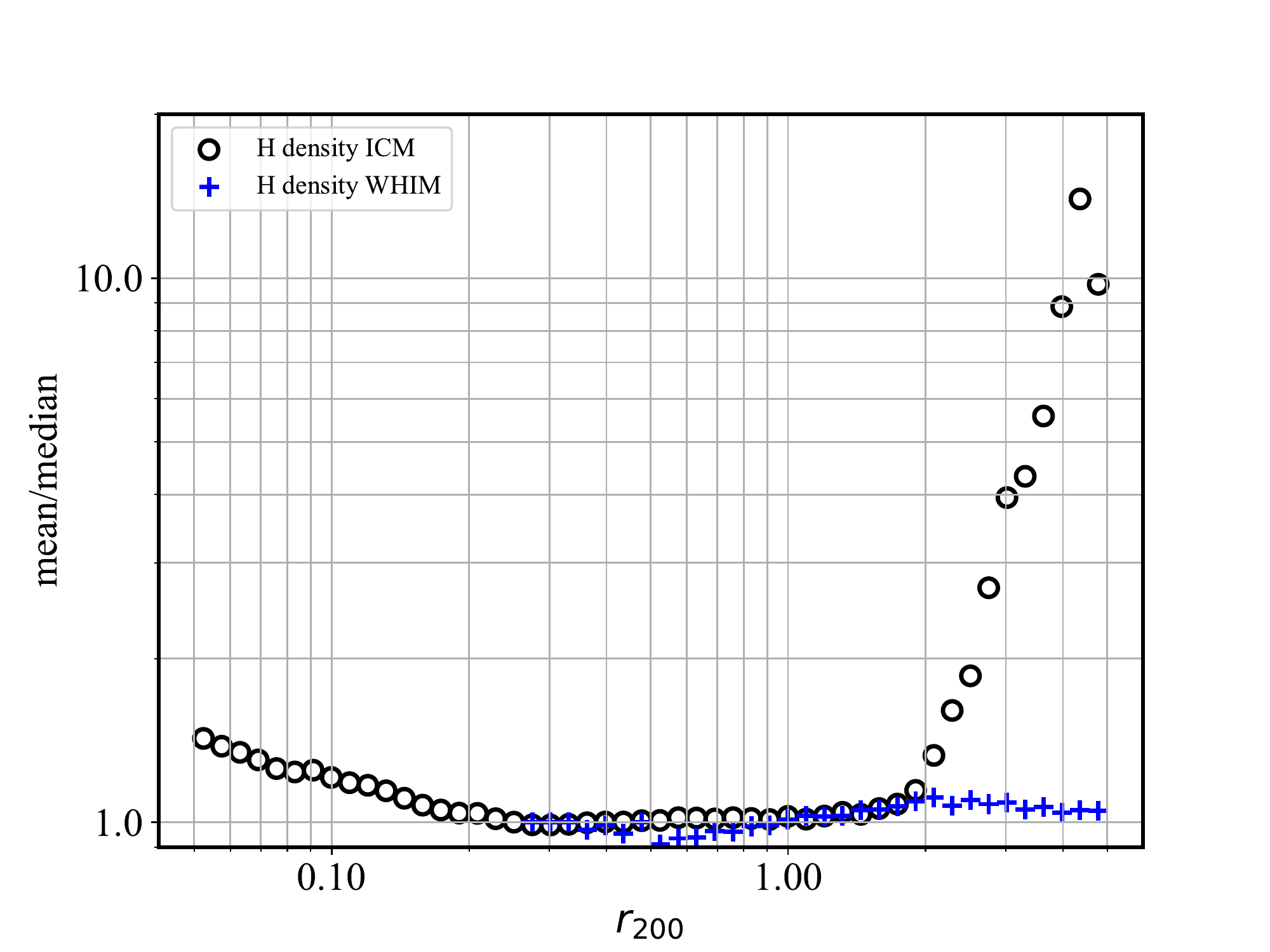}
    \includegraphics[width=2.8in]{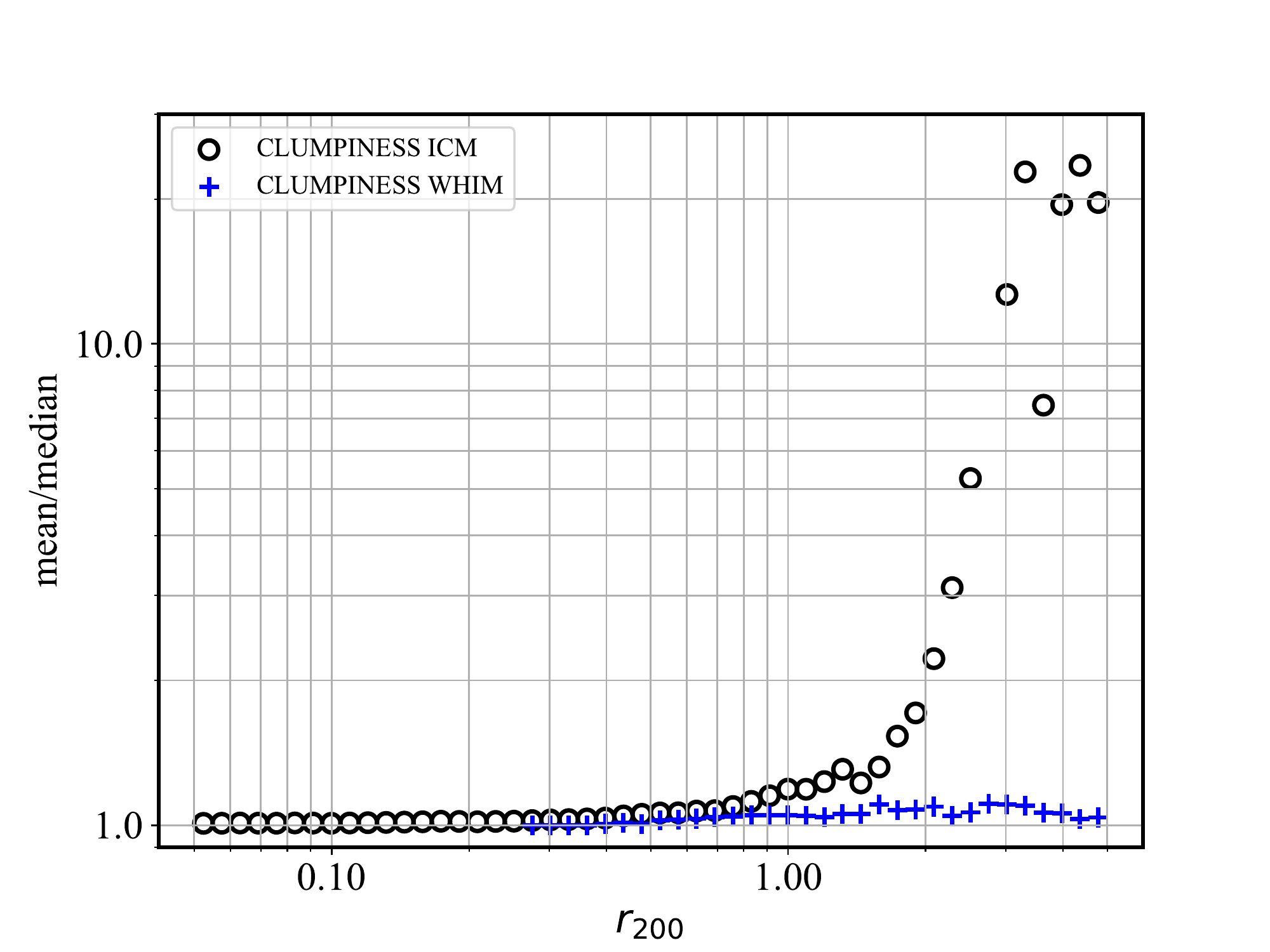}
    \caption{Radial distribution of the ratio of the mean to the median for selected quantities. 
        Top: $n_H$ density for the ICM and the WARM gas; Bottom: clumpiness factor.}
    \label{fig:meanvsmedian}
\end{figure}

With the distribution of the value of the quantity for the $\leq$~\nclu\ that have data at that
radius, we can then calculate key statistics such as the mean, median and standard deviation.
In Fig.~\ref{fig:distr} we show selected distributions to illustrate the method of analysis.
For example, typical distributions of the ICM density use all of the \nclu\ clusters as in the top panel of
Fig.~\ref{fig:distr}, while often only a smaller number of clusters have WHIM or WARM particles
at small distances from the cluster center. The bottom panel of Fig.~\ref{fig:distr} illustrates
a typical situation with the distribution of the clumpiness factor $C$ for the WARM and WHIM phases,
whereas the distribution is significantly skewed, resulting in a large difference between the mean and the
median. Radial profiles of the ratio of the mean to the median of the distributions, for the ICM and WARM $n_H$ density and for the 
clumpiness factor, are shown in Fig.~\ref{fig:meanvsmedian}, to illustrate the difference between 
the two statistics. We choose to report the median as the most representative value of the distribution, and its standard deviation as obtained from a Monte Carlo bootstrap resampling method, as a measure of the variability in the distribution. 

The choice to use the median as a representative value for a parameter is well documented
in the statistical literature, because of its greater insensitivity to outliers, compared to the 
sample mean \citep[for a review, see, e.g.][]{bonamente2022book}.  In fact, most studies of the
clumpiness of the gas use the median as a representative value \citep[e.g.][]{Planelles2017,nagai2011},
and similarly for other thermodynamic quantities \citep[e.g.][]{zhuravleva2013,eckert2015}. \cite{Towler2023} also investigates the difference between the mean and median of certain
thermodynamical quantities in simulations (see their Fig.~6), and conclude that mean of density--related quantities
are indeed more susceptible to bias due to effect of substructures, compared to the median.
In our simulations, we also found this to be true; in fact, in the top panel of Fig.~\ref{fig:meanvsmedian}
the datapoint corresponding to $r/r_{200}=2.75$ is not plotted, as it falls significantly above the scale (for a mean/median ratio above 10,000) due to the effect of a small clump of HOT gas near that radius that is
likely unrelated to the hot ICM of one cluster in the sample (cluster number 130).

\section{Tables of radial evolution of cluster properties}
\label{Appendix:Tables}
 This appendix reports the average radial profiles of 
 the thermodynamic properties of gas, namely density and 
 volume covering fraction in Table~\ref{tab:gasProperties}, 
 temperature and metal abundance in Table~\ref{tab:gasProperties2}, and clumping factor in Table~\ref{tab:gasProperties3}.
 Each average quantity is reported as a function of phase (ICM, WHIM and WARM) and for selected radial distances from the cluster's center. In each table, the first block is for the median quantity of all the clusters, 
 and the other blocks respectively for the three mass bins in order
 of increasing average mass ($M_1$, $M_2$, and $M_3$).
 
    \begin{table*}
        \footnotesize
        \centering
        \caption{Density and volume covering fractions of the ICM, WHIM and WARM gas phases in the \tng\ clusters, for representative radii (see Fig.~\ref{fig:ElDens} and \ref{fig:V}). The four blocks correspond respectively to all clusters followed by
        mass samples $M_1$, $M_2$ and $M_3$. }
        \begin{adjustbox}{angle=-90}
        \begin{tabular}{l|llllll}
        \hline 
        \hline
         & \multicolumn{3}{c}{Average $n_e$ Density (cm$^{3}$) from \eqref{eq:Qaverage}} & \multicolumn{3}{c}{Volume covering fraction $f_k$ from \eqref{eq:f}} \\
         & \multicolumn{3}{c}{ \hrulefill} & \multicolumn{3}{c}{ \hrulefill}  \\
          $r/r_{200}$  & ICM & WHIM & WARM & ICM & WHIM & WARM \\
        \hline
        0.10 &$ 1.97 \pm 0.08 \times 10^{-3} $&  ... & $ 9.46 \pm 1.79 \times 10^{-3} $& $ 1.00 \pm 0.00$&  ... & $ 5.59 \pm 1.72 \times 10^{-4} $\\
0.30 &$ 5.83 \pm 0.07 \times 10^{-4} $&  ... & $ 1.68 \pm 0.13 \times 10^{-3} $& $ 1.00 \pm 0.00$&  ... & $ 2.04 \pm 0.35 \times 10^{-4} $\\
0.50 &$ 2.31 \pm 0.02 \times 10^{-4} $& $ 8.78 \pm 0.13 \times 10^{-5} $& $ 4.23 \pm 0.18 \times 10^{-4} $& $ 1.00 \pm 0.00$& $ 4.58 \pm 2.54 \times 10^{-3} $& $ 1.47 \pm 0.33 \times 10^{-3} $\\
0.70 &$ 9.75 \pm 0.07 \times 10^{-5} $& $ 8.14 \pm 0.12 \times 10^{-5} $& $ 1.19 \pm 0.03 \times 10^{-4} $& $ 0.97 \pm 0.01$& $ 2.23 \pm 0.45 \times 10^{-2} $& $ 3.29 \pm 0.78 \times 10^{-2} $\\
1.00 &$ 3.49 \pm 0.07 \times 10^{-5} $& $ 3.42 \pm 0.06 \times 10^{-5} $& $ 3.56 \pm 0.10 \times 10^{-5} $& $ 0.69 \pm 0.02$& $ 0.30 \pm 0.02$& $ 0.31 \pm 0.02$\\
2.00 &$ 2.27 \pm 0.07 \times 10^{-6} $& $ 3.30 \pm 0.07 \times 10^{-6} $& $ 3.34 \pm 0.08 \times 10^{-6} $& $ 0.17 \pm 0.02$& $ 0.83 \pm 0.02$& $ 0.84 \pm 0.02$\\
3.00 &$ 4.63 \pm 0.21 \times 10^{-7} $& $ 1.49 \pm 0.03 \times 10^{-6} $& $ 1.52 \pm 0.07 \times 10^{-6} $& $ 0.11 \pm 0.01$& $ 0.83 \pm 0.01$& $ 0.82 \pm 0.01$\\
4.00 &$ 2.35 \pm 0.29 \times 10^{-7} $& $ 1.19 \pm 0.03 \times 10^{-6} $& $ 1.22 \pm 0.04 \times 10^{-6} $& $ 5.86 \pm 0.69 \times 10^{-2} $& $ 0.63 \pm 0.01$& $ 0.63 \pm 0.01$\\
5.00 &$ 3.83 \pm 0.72 \times 10^{-7} $& $ 1.22 \pm 0.05 \times 10^{-6} $& $ 1.24 \pm 0.05 \times 10^{-6} $& $ 1.67 \pm 0.23 \times 10^{-2} $& $ 0.44 \pm 0.01$& $ 0.44 \pm 0.01$\\
\hline
0.10 &$ 1.82 \pm 0.06 \times 10^{-3} $&  ... & $ 7.68 \pm 0.94 \times 10^{-3} $& $ 1.00 \pm 0.00$&  ... & $ 8.46 \pm 3.87 \times 10^{-4} $\\
0.30 &$ 5.71 \pm 0.18 \times 10^{-4} $&  ... & $ 1.51 \pm 0.11 \times 10^{-3} $& $ 1.00 \pm 0.00$&  ... & $ 4.87 \pm 1.85 \times 10^{-4} $\\
0.50 &$ 2.35 \pm 0.03 \times 10^{-4} $& $ 9.20 \pm 0.97 \times 10^{-5} $& $ 3.64 \pm 0.34 \times 10^{-4} $& $ 1.00 \pm 0.00$& $ 5.17 \pm 3.06 \times 10^{-3} $& $ 3.08 \pm 0.43 \times 10^{-3} $\\
0.70 &$ 9.80 \pm 0.07 \times 10^{-5} $& $ 8.48 \pm 0.16 \times 10^{-5} $& $ 1.05 \pm 0.03 \times 10^{-4} $& $ 0.91 \pm 0.01$& $ 5.92 \pm 1.37 \times 10^{-2} $& $ 9.25 \pm 2.66 \times 10^{-2} $\\
1.00 &$ 3.43 \pm 0.05 \times 10^{-5} $& $ 3.47 \pm 0.07 \times 10^{-5} $& $ 3.57 \pm 0.10 \times 10^{-5} $& $ 0.52 \pm 0.01$& $ 0.47 \pm 0.02$& $ 0.48 \pm 0.02$\\
2.00 &$ 2.19 \pm 0.31 \times 10^{-6} $& $ 3.16 \pm 0.09 \times 10^{-6} $& $ 3.20 \pm 0.07 \times 10^{-6} $& $ 6.30 \pm 1.15 \times 10^{-2} $& $ 0.94 \pm 0.01$& $ 0.94 \pm 0.01$\\
3.00 &$ 4.46 \pm 0.63 \times 10^{-7} $& $ 1.38 \pm 0.06 \times 10^{-6} $& $ 1.43 \pm 0.07 \times 10^{-6} $& $ 4.31 \pm 1.00 \times 10^{-2} $& $ 0.92 \pm 0.01$& $ 0.92 \pm 0.01$\\
4.00 &$ 2.09 \pm 0.27 \times 10^{-7} $& $ 1.07 \pm 0.06 \times 10^{-6} $& $ 1.07 \pm 0.05 \times 10^{-6} $& $ 2.50 \pm 0.68 \times 10^{-2} $& $ 0.72 \pm 0.01$& $ 0.73 \pm 0.01$\\
5.00 &$ 3.08 \pm 0.82 \times 10^{-7} $& $ 1.06 \pm 0.09 \times 10^{-6} $& $ 1.08 \pm 0.09 \times 10^{-6} $& $ 8.78 \pm 3.38 \times 10^{-3} $& $ 0.50 \pm 0.01$& $ 0.50 \pm 0.01$\\
\hline
0.10 &$ 1.99 \pm 0.08 \times 10^{-3} $&  ... & $ 1.37 \pm 0.30 \times 10^{-2} $& $ 1.00 \pm 0.00$&  ... & $ 2.86 \pm 1.12 \times 10^{-4} $\\
0.30 &$ 5.83 \pm 0.06 \times 10^{-4} $&  ... & $ 1.78 \pm 0.19 \times 10^{-3} $& $ 1.00 \pm 0.00$&  ... & $ 1.28 \pm 0.36 \times 10^{-4} $\\
0.50 &$ 2.29 \pm 0.02 \times 10^{-4} $& $ 8.72 \pm 0.27 \times 10^{-5} $& $ 4.62 \pm 0.54 \times 10^{-4} $& $ 1.00 \pm 0.00$& $ 7.17 \pm 5.82 \times 10^{-3} $& $ 7.43 \pm 1.82 \times 10^{-4} $\\
0.70 &$ 9.80 \pm 0.12 \times 10^{-5} $& $ 7.97 \pm 0.13 \times 10^{-5} $& $ 1.28 \pm 0.04 \times 10^{-4} $& $ 0.99 \pm 0.00$& $ 8.66 \pm 4.01 \times 10^{-3} $& $ 1.44 \pm 0.36 \times 10^{-2} $\\
1.00 &$ 3.59 \pm 0.09 \times 10^{-5} $& $ 3.35 \pm 0.07 \times 10^{-5} $& $ 3.43 \pm 0.07 \times 10^{-5} $& $ 0.80 \pm 0.03$& $ 0.21 \pm 0.04$& $ 0.21 \pm 0.04$\\
2.00 &$ 2.40 \pm 0.12 \times 10^{-6} $& $ 3.36 \pm 0.12 \times 10^{-6} $& $ 3.41 \pm 0.16 \times 10^{-6} $& $ 0.21 \pm 0.01$& $ 0.78 \pm 0.02$& $ 0.79 \pm 0.01$\\
3.00 &$ 4.89 \pm 0.35 \times 10^{-7} $& $ 1.52 \pm 0.08 \times 10^{-6} $& $ 1.56 \pm 0.08 \times 10^{-6} $& $ 0.16 \pm 0.01$& $ 0.79 \pm 0.02$& $ 0.79 \pm 0.01$\\
4.00 &$ 3.03 \pm 0.65 \times 10^{-7} $& $ 1.23 \pm 0.06 \times 10^{-6} $& $ 1.25 \pm 0.04 \times 10^{-6} $& $ 7.02 \pm 0.92 \times 10^{-2} $& $ 0.61 \pm 0.01$& $ 0.61 \pm 0.02$\\
5.00 &$ 3.89 \pm 0.50 \times 10^{-7} $& $ 1.26 \pm 0.05 \times 10^{-6} $& $ 1.30 \pm 0.04 \times 10^{-6} $& $ 2.17 \pm 0.63 \times 10^{-2} $& $ 0.42 \pm 0.01$& $ 0.42 \pm 0.01$\\
\hline
0.10 &$ 3.41 \pm 0.28 \times 10^{-3} $&  ... &  ... & $ 1.00 \pm 0.00$&  ... &  ... \\
0.30 &$ 6.35 \pm 0.36 \times 10^{-4} $&  ... & $ 4.80 \pm 0.54 \times 10^{-3} $& $ 1.00 \pm 0.00$&  ... & $ 1.82 \pm 0.55 \times 10^{-4} $\\
0.50 &$ 2.32 \pm 0.03 \times 10^{-4} $& $ 8.24 \pm 0.83 \times 10^{-5} $& $ 1.63 \pm 0.29 \times 10^{-3} $& $ 1.00 \pm 0.00$& $ 7.30 \pm 44.55 \times 10^{-4} $& $ 3.84 \pm 0.32 \times 10^{-4} $\\
0.70 &$ 9.10 \pm 0.48 \times 10^{-5} $& $ 7.45 \pm 0.30 \times 10^{-5} $& $ 2.31 \pm 0.77 \times 10^{-4} $& $ 1.00 \pm 0.00$& $ 1.51 \pm 0.33 \times 10^{-3} $& $ 9.98 \pm 5.50 \times 10^{-4} $\\
1.00 &$ 3.16 \pm 0.15 \times 10^{-5} $& $ 3.68 \pm 0.10 \times 10^{-5} $& $ 4.95 \pm 0.67 \times 10^{-5} $& $ 0.99 \pm 0.00$& $ 1.48 \pm 0.64 \times 10^{-2} $& $ 1.77 \pm 0.66 \times 10^{-2} $\\
2.00 &$ 1.87 \pm 0.20 \times 10^{-6} $& $ 4.11 \pm 0.67 \times 10^{-6} $& $ 4.23 \pm 0.79 \times 10^{-6} $& $ 0.58 \pm 0.01$& $ 0.42 \pm 0.02$& $ 0.42 \pm 0.01$\\
3.00 &$ 4.68 \pm 0.21 \times 10^{-7} $& $ 1.75 \pm 0.08 \times 10^{-6} $& $ 1.75 \pm 0.24 \times 10^{-6} $& $ 0.37 \pm 0.09$& $ 0.55 \pm 0.02$& $ 0.56 \pm 0.02$\\
4.00 &$ 2.51 \pm 0.27 \times 10^{-7} $& $ 1.29 \pm 0.05 \times 10^{-6} $& $ 1.31 \pm 0.05 \times 10^{-6} $& $ 0.18 \pm 0.07$& $ 0.45 \pm 0.02$& $ 0.45 \pm 0.02$\\
5.00 &$ 1.18 \pm 0.24 \times 10^{-6} $& $ 1.49 \pm 0.13 \times 10^{-6} $& $ 1.56 \pm 0.04 \times 10^{-6} $& $ 4.80 \pm 2.31 \times 10^{-2} $& $ 0.34 \pm 0.01$& $ 0.34 \pm 0.01$ \\
\hline
\hline
        \end{tabular}
        \end{adjustbox}
        \label{tab:gasProperties}
    \end{table*}

\begin{table*}
\footnotesize
    \centering
        \caption{Temperature and chemical abundance of the ICM, WHIM and WARM gas phases in the \tng\ clusters, for representative radii (see Fig.~\ref{fig:T} and \ref{fig:A}). The four blocks correspond respectively to all clusters followed by
        mass samples $M_1$, $M_2$ and $M_3$. }
    \setlength\extrarowheight{-1pt}
    \begin{adjustbox}{angle=-90}   
    \begin{tabular}{l|llllll}
    \hline 
    \hline
     & \multicolumn{3}{c}{Average $T$ (K)}& \multicolumn{3}{c}{Average metal abundance A (Solar)} \\
     & \multicolumn{3}{c}{ \hrulefill} & \multicolumn{3}{c}{ \hrulefill}  \\
      $r/r_{200}$  & ICM & WHIM & WARM & ICM & WHIM & WARM \\
    \hline
    0.10 &$ 3.54 \pm 0.10 \times 10^{7} $&  ... & $ 6.52 \pm 0.29 \times 10^{6} $& $ 0.29 \pm 0.01$&  ... & $ 0.68 \pm 0.06$\\
0.30 &$ 2.46 \pm 0.05 \times 10^{7} $&  ... & $ 8.09 \pm 0.09 \times 10^{6} $& $ 0.22 \pm 0.00$&  ... & $ 0.45 \pm 0.01$\\
0.50 &$ 2.06 \pm 0.05 \times 10^{7} $& $ 8.74 \pm 0.33 \times 10^{6} $& $ 8.44 \pm 0.12 \times 10^{6} $& $ 0.18 \pm 0.00$& $ 0.16 \pm 0.01$& $ 0.30 \pm 0.01$\\
0.70 &$ 1.77 \pm 0.07 \times 10^{7} $& $ 8.94 \pm 0.06 \times 10^{6} $& $ 8.67 \pm 0.03 \times 10^{6} $& $ 0.14 \pm 0.00$& $ 0.17 \pm 0.00$& $ 0.20 \pm 0.00$\\
1.00 &$ 1.50 \pm 0.03 \times 10^{7} $& $ 7.98 \pm 0.03 \times 10^{6} $& $ 7.98 \pm 0.03 \times 10^{6} $& $ 0.11 \pm 0.00$& $ 0.14 \pm 0.00$& $ 0.14 \pm 0.00$\\
2.00 &$ 1.35 \pm 0.01 \times 10^{7} $& $ 4.99 \pm 0.11 \times 10^{6} $& $ 4.99 \pm 0.12 \times 10^{6} $& $ 5.13 \pm 0.19 \times 10^{-2} $& $ 8.30 \pm 0.13 \times 10^{-2} $& $ 8.31 \pm 0.13 \times 10^{-2} $\\
3.00 &$ 1.39 \pm 0.03 \times 10^{7} $& $ 3.92 \pm 0.05 \times 10^{6} $& $ 3.91 \pm 0.06 \times 10^{6} $& $ 2.27 \pm 0.19 \times 10^{-2} $& $ 5.50 \pm 0.17 \times 10^{-2} $& $ 5.53 \pm 0.18 \times 10^{-2} $\\
4.00 &$ 1.40 \pm 0.02 \times 10^{7} $& $ 2.89 \pm 0.11 \times 10^{6} $& $ 2.90 \pm 0.10 \times 10^{6} $& $ 9.49 \pm 0.52 \times 10^{-3} $& $ 4.60 \pm 0.22 \times 10^{-2} $& $ 4.64 \pm 0.23 \times 10^{-2} $\\
5.00 &$ 1.34 \pm 0.01 \times 10^{7} $& $ 2.19 \pm 0.07 \times 10^{6} $& $ 2.20 \pm 0.09 \times 10^{6} $& $ 9.76 \pm 0.89 \times 10^{-3} $& $ 4.95 \pm 0.17 \times 10^{-2} $& $ 4.91 \pm 0.20 \times 10^{-2} $\\
\hline
0.10 &$ 3.20 \pm 0.03 \times 10^{7} $&  ... & $ 6.56 \pm 0.20 \times 10^{6} $& $ 0.30 \pm 0.01$&  ... & $ 0.69 \pm 0.06$\\
0.30 &$ 2.09 \pm 0.01 \times 10^{7} $&  ... & $ 7.93 \pm 0.36 \times 10^{6} $& $ 0.24 \pm 0.00$&  ... & $ 0.38 \pm 0.06$\\
0.50 &$ 1.70 \pm 0.01 \times 10^{7} $& $ 9.13 \pm 0.24 \times 10^{6} $& $ 8.74 \pm 0.18 \times 10^{6} $& $ 0.19 \pm 0.00$& $ 0.16 \pm 0.01$& $ 0.28 \pm 0.01$\\
0.70 &$ 1.46 \pm 0.02 \times 10^{7} $& $ 9.07 \pm 0.11 \times 10^{6} $& $ 8.88 \pm 0.14 \times 10^{6} $& $ 0.14 \pm 0.00$& $ 0.18 \pm 0.00$& $ 0.20 \pm 0.00$\\
1.00 &$ 1.31 \pm 0.02 \times 10^{7} $& $ 7.96 \pm 0.03 \times 10^{6} $& $ 7.95 \pm 0.02 \times 10^{6} $& $ 0.10 \pm 0.00$& $ 0.14 \pm 0.01$& $ 0.14 \pm 0.01$\\
2.00 &$ 1.24 \pm 0.01 \times 10^{7} $& $ 4.51 \pm 0.12 \times 10^{6} $& $ 4.54 \pm 0.15 \times 10^{6} $& $ 4.44 \pm 0.16 \times 10^{-2} $& $ 8.26 \pm 0.33 \times 10^{-2} $& $ 8.33 \pm 0.32 \times 10^{-2} $\\
3.00 &$ 1.25 \pm 0.04 \times 10^{7} $& $ 3.62 \pm 0.15 \times 10^{6} $& $ 3.62 \pm 0.14 \times 10^{6} $& $ 1.88 \pm 0.18 \times 10^{-2} $& $ 5.59 \pm 0.34 \times 10^{-2} $& $ 5.53 \pm 0.22 \times 10^{-2} $\\
4.00 &$ 1.24 \pm 0.02 \times 10^{7} $& $ 2.79 \pm 0.15 \times 10^{6} $& $ 2.78 \pm 0.14 \times 10^{6} $& $ 6.34 \pm 1.05 \times 10^{-3} $& $ 4.45 \pm 0.30 \times 10^{-2} $& $ 4.41 \pm 0.30 \times 10^{-2} $\\
5.00 &$ 1.25 \pm 0.03 \times 10^{7} $& $ 1.97 \pm 0.28 \times 10^{6} $& $ 2.01 \pm 0.22 \times 10^{6} $& $ 7.33 \pm 1.74 \times 10^{-3} $& $ 4.73 \pm 0.22 \times 10^{-2} $& $ 4.72 \pm 0.22 \times 10^{-2} $\\
\hline
0.10 &$ 3.74 \pm 0.05 \times 10^{7} $&  ... & $ 6.45 \pm 0.81 \times 10^{6} $& $ 0.29 \pm 0.01$&  ... & $ 0.64 \pm 0.12$\\
0.30 &$ 2.66 \pm 0.07 \times 10^{7} $&  ... & $ 8.10 \pm 0.14 \times 10^{6} $& $ 0.22 \pm 0.00$&  ... & $ 0.47 \pm 0.04$\\
0.50 &$ 2.27 \pm 0.09 \times 10^{7} $& $ 8.42 \pm 0.49 \times 10^{6} $& $ 8.37 \pm 0.14 \times 10^{6} $& $ 0.17 \pm 0.00$& $ 0.17 \pm 0.02$& $ 0.31 \pm 0.02$\\
0.70 &$ 1.97 \pm 0.04 \times 10^{7} $& $ 8.88 \pm 0.14 \times 10^{6} $& $ 8.51 \pm 0.08 \times 10^{6} $& $ 0.14 \pm 0.00$& $ 0.17 \pm 0.01$& $ 0.21 \pm 0.01$\\
1.00 &$ 1.62 \pm 0.04 \times 10^{7} $& $ 8.16 \pm 0.09 \times 10^{6} $& $ 8.14 \pm 0.09 \times 10^{6} $& $ 0.11 \pm 0.00$& $ 0.14 \pm 0.01$& $ 0.14 \pm 0.01$\\
2.00 &$ 1.41 \pm 0.03 \times 10^{7} $& $ 5.24 \pm 0.16 \times 10^{6} $& $ 5.21 \pm 0.16 \times 10^{6} $& $ 5.47 \pm 0.28 \times 10^{-2} $& $ 8.37 \pm 0.16 \times 10^{-2} $& $ 8.33 \pm 0.12 \times 10^{-2} $\\
3.00 &$ 1.48 \pm 0.02 \times 10^{7} $& $ 4.04 \pm 0.11 \times 10^{6} $& $ 4.02 \pm 0.11 \times 10^{6} $& $ 2.59 \pm 0.21 \times 10^{-2} $& $ 5.49 \pm 0.20 \times 10^{-2} $& $ 5.52 \pm 0.16 \times 10^{-2} $\\
4.00 &$ 1.45 \pm 0.03 \times 10^{7} $& $ 2.99 \pm 0.15 \times 10^{6} $& $ 2.98 \pm 0.16 \times 10^{6} $& $ 1.05 \pm 0.09 \times 10^{-2} $& $ 4.70 \pm 0.25 \times 10^{-2} $& $ 4.75 \pm 0.27 \times 10^{-2} $\\
5.00 &$ 1.36 \pm 0.01 \times 10^{7} $& $ 2.29 \pm 0.11 \times 10^{6} $& $ 2.29 \pm 0.11 \times 10^{6} $& $ 1.06 \pm 0.08 \times 10^{-2} $& $ 5.16 \pm 0.42 \times 10^{-2} $& $ 5.19 \pm 0.43 \times 10^{-2} $\\
\hline
0.10 &$ 6.11 \pm 0.65 \times 10^{7} $&  ... &  ... & $ 0.25 \pm 0.01$&  ... &  ... \\
0.30 &$ 4.82 \pm 0.67 \times 10^{7} $&  ... & $ 7.02 \pm 0.87 \times 10^{6} $& $ 0.19 \pm 0.00$&  ... & $ 0.90 \pm 0.33$\\
0.50 &$ 4.24 \pm 0.07 \times 10^{7} $& $ 8.42 \pm 0.26 \times 10^{6} $& $ 6.87 \pm 0.29 \times 10^{6} $& $ 0.16 \pm 0.00$& $ 0.14 \pm 0.04$& $ 0.36 \pm 0.08$\\
0.70 &$ 3.43 \pm 0.36 \times 10^{7} $& $ 8.43 \pm 0.18 \times 10^{6} $& $ 7.86 \pm 0.18 \times 10^{6} $& $ 0.13 \pm 0.00$& $ 0.14 \pm 0.02$& $ 0.21 \pm 0.05$\\
1.00 &$ 2.58 \pm 0.11 \times 10^{7} $& $ 7.66 \pm 0.18 \times 10^{6} $& $ 7.54 \pm 0.12 \times 10^{6} $& $ 0.11 \pm 0.00$& $ 0.14 \pm 0.01$& $ 0.14 \pm 0.01$\\
2.00 &$ 1.79 \pm 0.10 \times 10^{7} $& $ 5.99 \pm 0.06 \times 10^{6} $& $ 6.02 \pm 0.05 \times 10^{6} $& $ 5.16 \pm 0.47 \times 10^{-2} $& $ 7.85 \pm 0.65 \times 10^{-2} $& $ 7.79 \pm 0.31 \times 10^{-2} $\\
3.00 &$ 1.75 \pm 0.05 \times 10^{7} $& $ 4.52 \pm 0.24 \times 10^{6} $& $ 4.50 \pm 0.41 \times 10^{6} $& $ 2.35 \pm 0.52 \times 10^{-2} $& $ 5.59 \pm 0.49 \times 10^{-2} $& $ 5.52 \pm 0.58 \times 10^{-2} $\\
4.00 &$ 1.87 \pm 0.21 \times 10^{7} $& $ 2.96 \pm 0.53 \times 10^{6} $& $ 3.05 \pm 0.38 \times 10^{6} $& $ 1.32 \pm 0.03 \times 10^{-2} $& $ 4.86 \pm 0.30 \times 10^{-2} $& $ 4.81 \pm 0.53 \times 10^{-2} $\\
5.00 &$ 1.76 \pm 0.05 \times 10^{7} $& $ 2.09 \pm 0.45 \times 10^{6} $& $ 2.12 \pm 0.45 \times 10^{6} $& $ 1.34 \pm 0.35 \times 10^{-2} $& $ 5.21 \pm 0.88 \times 10^{-2} $& $ 5.20 \pm 0.87 \times 10^{-2} $ \\
\hline
\hline
    \end{tabular}
    \end{adjustbox}
    \label{tab:gasProperties2}
\end{table*}

\begin{table*}
\footnotesize
    \centering
        \caption{Clumping factor of the ICM, WHIM and WARM gas phases in the \tng\ clusters, for representative radii (see Fig.~\ref{fig:clumping}), and for all phases combined. The four blocks correspond respectively to all clusters, followed by
        mass samples $M_1$, $M_2$ and $M_3$.} 
    \setlength\extrarowheight{0pt}
    \begin{tabular}{l|llll}
    \hline 
    \hline
     & \multicolumn{4}{c}{Density clumping factor $C$ from Eq.~\eqref{eq:C}} \\
     & \multicolumn{4}{c}{ \hrulefill} \\
      $r/r_{200}$  & ICM & WHIM & WARM & All phases\\
    \hline
    0.10 &$ 1.03 \pm 0.00$&  \nodata & $ 1.32 \pm 0.07$& $ 1.03 \pm 0.00$\\
0.30 &$ 1.07 \pm 0.00$&  \nodata & $ 1.31 \pm 0.04$& $ 1.08 \pm 0.01$\\
0.50 &$ 1.10 \pm 0.00$& $ 1.03 \pm 0.01$& $ 1.68 \pm 0.09$& $ 1.14 \pm 0.01$\\
0.70 &$ 1.13 \pm 0.01$& $ 1.05 \pm 0.01$& $ 1.56 \pm 0.10$& $ 1.34 \pm 0.09$\\
1.00 &$ 1.20 \pm 0.02$& $ 1.21 \pm 0.01$& $ 1.50 \pm 0.06$& $ 1.78 \pm 0.14$\\
2.00 &$ 1.93 \pm 0.07$& $ 2.91 \pm 0.08$& $ 4.29 \pm 0.25$& $ 34.75 \pm 4.82$\\
3.00 &$ 5.63 \pm 1.03$& $ 6.75 \pm 0.22$& $ 11.73 \pm 0.66$& $ 2.46 \pm 0.36 \times 10^{2} $\\
4.00 &$ 13.29 \pm 2.08$& $ 9.50 \pm 0.26$& $ 17.00 \pm 0.92$& $ 5.41 \pm 0.49 \times 10^{2} $\\
5.00 &$ 31.29 \pm 6.64$& $ 10.09 \pm 0.26$& $ 20.52 \pm 1.14$& $ 8.80 \pm 1.02 \times 10^{2} $\\
\hline
0.10 &$ 1.02 \pm 0.00$&  \nodata & $ 1.36 \pm 0.07$& $ 1.02 \pm 0.01$\\
0.30 &$ 1.05 \pm 0.00$&  \nodata & $ 1.24 \pm 0.13$& $ 1.05 \pm 0.01$\\
0.50 &$ 1.08 \pm 0.00$& $ 1.03 \pm 0.02$& $ 1.64 \pm 0.16$& $ 1.12 \pm 0.02$\\
0.70 &$ 1.10 \pm 0.00$& $ 1.04 \pm 0.01$& $ 1.28 \pm 0.05$& $ 1.25 \pm 0.06$\\
1.00 &$ 1.17 \pm 0.01$& $ 1.17 \pm 0.01$& $ 1.32 \pm 0.04$& $ 1.65 \pm 0.15$\\
2.00 &$ 1.80 \pm 0.09$& $ 2.74 \pm 0.08$& $ 3.39 \pm 0.35$& $ 26.44 \pm 9.91$\\
3.00 &$ 3.71 \pm 0.53$& $ 7.05 \pm 0.31$& $ 11.85 \pm 1.01$& $ 3.14 \pm 0.75 \times 10^{2} $\\
4.00 &$ 9.57 \pm 2.81$& $ 9.43 \pm 0.90$& $ 14.66 \pm 0.65$& $ 4.54 \pm 1.35 \times 10^{2} $\\
5.00 &$ 26.99 \pm 9.98$& $ 10.18 \pm 0.63$& $ 18.86 \pm 4.06$& $ 7.66 \pm 1.70 \times 10^{2} $\\
\hline
0.10 &$ 1.03 \pm 0.00$&  \nodata & $ 1.29 \pm 0.02$& $ 1.03 \pm 0.00$\\
0.30 &$ 1.08 \pm 0.01$&  \nodata & $ 1.32 \pm 0.03$& $ 1.09 \pm 0.01$\\
0.50 &$ 1.10 \pm 0.01$& $ 1.03 \pm 0.01$& $ 1.60 \pm 0.05$& $ 1.14 \pm 0.01$\\
0.70 &$ 1.16 \pm 0.01$& $ 1.06 \pm 0.01$& $ 1.79 \pm 0.11$& $ 1.43 \pm 0.14$\\
1.00 &$ 1.23 \pm 0.03$& $ 1.22 \pm 0.01$& $ 1.55 \pm 0.05$& $ 1.69 \pm 0.09$\\
2.00 &$ 2.00 \pm 0.08$& $ 3.07 \pm 0.19$& $ 4.47 \pm 0.29$& $ 31.12 \pm 5.05$\\
3.00 &$ 7.61 \pm 1.22$& $ 6.71 \pm 0.18$& $ 10.84 \pm 1.03$& $ 1.97 \pm 0.33 \times 10^{2} $\\
4.00 &$ 15.67 \pm 3.78$& $ 9.70 \pm 0.38$& $ 21.01 \pm 3.67$& $ 5.77 \pm 1.36 \times 10^{2} $\\
5.00 &$ 26.90 \pm 7.74$& $ 10.01 \pm 0.27$& $ 20.45 \pm 0.69$& $ 8.71 \pm 0.99 \times 10^{2} $\\
\hline
0.10 &$ 1.07 \pm 0.02$&  \nodata &  \nodata & $ 1.07 \pm 0.01$\\
0.30 &$ 1.10 \pm 0.02$&  \nodata & $ 1.33 \pm 0.21$& $ 1.11 \pm 0.03$\\
0.50 &$ 1.16 \pm 0.02$& $ 1.03 \pm 0.01$& $ 2.72 \pm 0.69$& $ 1.25 \pm 0.07$\\
0.70 &$ 1.23 \pm 0.07$& $ 1.09 \pm 0.03$& $ 2.37 \pm 0.49$& $ 1.55 \pm 0.18$\\
1.00 &$ 1.24 \pm 0.03$& $ 1.41 \pm 0.03$& $ 4.07 \pm 0.25$& $ 9.10 \pm 2.22$\\
2.00 &$ 2.14 \pm 0.61$& $ 3.29 \pm 0.11$& $ 5.83 \pm 0.45$& $ 1.10 \pm 0.41 \times 10^{2} $\\
3.00 &$ 10.84 \pm 2.06$& $ 6.29 \pm 0.36$& $ 16.16 \pm 3.43$& $ 6.48 \pm 2.07 \times 10^{2} $\\
4.00 &$ 37.51 \pm 31.92$& $ 9.14 \pm 0.61$& $ 18.98 \pm 1.11$& $ 8.34 \pm 2.63 \times 10^{2} $\\
5.00 &$ 74.81 \pm 26.86$& $ 10.13 \pm 1.13$& $ 33.21 \pm 7.25$& $ 1.25 \pm 0.36 \times 10^{3} $\\
\hline
\hline
    \end{tabular}
    \label{tab:gasProperties3}
\end{table*}

\end{document}